\documentclass[submission,copyright,creativecommons]{eptcs}

\providecommand{\event}{FOCLASA 2011}

\usepackage{latexsym,amsmath,amssymb,amsfonts,stmaryrd}
\usepackage[mathscr]{eucal}
\usepackage{graphicx}
\usepackage{alltt}
\usepackage{url}
\usepackage{colortbl}
\usepackage{xspace}  
\usepackage{cancel}
\usepackage{semantic} 
\usepackage{color} 
\definecolor{darkblue}{rgb}{0,0,0.7} 
\usepackage{enumerate} 
\usepackage{longtable} 
\usepackage{booktabs} 
\usepackage{wrapfig} 
\usepackage{listings} 
\usepackage{ifthen}
\usepackage{todonotes} 
\usepackage{pdfsync} 
\usepackage{datetime} 
\usepackage{verbatim} 

\usepackage{tikz}
\usepgflibrary{decorations.pathreplacing}
\usetikzlibrary{%
  arrows,%
  calc,%
  fit,%
  shadows,%
  patterns,%
  matrix,%
  decorations.pathmorphing,%
  shapes,automata,decorations.markings
}
\pgfdeclarelayer{background}
\pgfdeclarelayer{threadground}
\pgfsetlayers{background,threadground,main}

\usepackage{src/extras/reo-tikz}     

\newcommand{\customref}[2]{\hyperref[#2]{#1~\ref*{#2}}}
\newcommand{\bref}[1]{\hyperref[#1]{(\ref*{#1})}}
\newcommand{\bcustomref}[2]{\hyperref[#2]{#1~(\ref*{#2})}}
\newcommand{\myref}[2]{%
  \ifthenelse{\equal{#1}{def}}{\customref{Definition}{#1:#2}}{%
  \ifthenelse{\equal{#1}{lem}}{\customref{Lemma}{#1:#2}}{%
  \ifthenelse{\equal{#1}{not}}{\customref{Notation}{#1:#2}}{%
  \ifthenelse{\equal{#1}{prop}}{\customref{Proposition}{#1:#2}}{%
  \ifthenelse{\equal{#1}{ax}}{\customref{Axiom}{#1:#2}}{%
  \ifthenelse{\equal{#1}{ex}}{\customref{Example}{#1:#2}}{%
  \ifthenelse{\equal{#1}{propr}}{\customref{Property}{#1:#2}}{%
  \ifthenelse{\equal{#1}{as}}{\customref{Assumption}{#1:#2}}{%
  \ifthenelse{\equal{#1}{tab}}{\customref{Table}{#1:#2}}{%
  \ifthenelse{\equal{#1}{sec}}{\hyperref[#1:#2]{\S\ref*{#1:#2}}}{%
  \ifthenelse{\equal{#1}{eq}}{\hyperref[#1:#2]{Equation~(\ref*{#1:#2})}}{%
  \autoref{#1:#2}%
  }}}}}}}}}}}}

\newtheorem{definition}{Definition}
\newtheorem{example}{Example}

\newtheorem{proposition}{Proposition}

\newcommand{\JP}[1]{\todo[color=blue!20]{J: #1}}

\newcommand{\JPin}[2][]{\todo[color=blue!20,inline, #1]{#2}}
\newcommand{\reviewin}[2][]{\todo[color=red!20,inline, #1]{Rev: #2}}

 \renewcommand{\todo}[2][~]{}

\newcommand{\myparagraph}[1]{\medskip\noindent\textbf{#1}~~}
\newcommand{\mytoprule}{\specialrule{.08em}{0pt}{0pt}\rule{0pt}{2.6ex}}
\newcommand{\mymidrule}{%
  \\[1mm]\hline\rule{0pt}{4mm}}
\newcommand{\mybottomrule}{%
  \\[1mm]\specialrule{.08em}{0pt}{0pt}}

\tikzstyle{stcloud} = [cloud,cloud puffs=7, draw, cloud ignores aspect,
                       minimum width=6mm]
\tikzstyle{st}   = [draw=black,minimum width=1mm,stcloud]
\tikzstyle{cdst} = [st,fill=black!30]
\tikzstyle{cgst} = [st,pattern=dots]
\tikzstyle{dtst} = [st,pattern=horizontal lines]
\tikzstyle{spst} = [st,draw=hidecolour,
                    pattern color=hidecolour,pattern=crosshatch]
\tikzstyle{prct} = [ultra thick]
\tikzstyle{arrw} = [ultra thick, decoration={markings,mark=at position
                    .85 with {\arrow[line width=1.3pt]{>}}}, 
                    postaction={decorate}]
\tikzstyle{ndistz} = [node distance=18mm]
\tikzstyle{ndistzz} = [node distance=9mm]
\tikzstyle{ndist} = [reodist] 
\tikzstyle{ndista} = [node distance=10mm]
\tikzstyle{ndistb} = [node distance=7.5mm]
\tikzstyle{ndistc} = [node distance=5mm]
\tikzstyle{ndistd} = [node distance=11mm]
\tikzstyle{prim}=[rectangle,rounded corners=2mm,minimum width=20mm,
                  inner sep=2mm,thick,draw=black,fill=white]
\tikzstyle{main}=[prim,ultra thick,top color=white, bottom color=blue!20]
\tikzstyle{mycloud}=[cloud,cloud puffs=11, draw, cloud ignores aspect]
\tikzstyle{observ}=[decorate,decoration={coil,aspect=0,segment length=2mm,  
                    amplitude=0.5mm},>=triangle 90]

\definecolor{bluetop}{rgb}{.85098039215686274509,.92941176470588235294,1}
\definecolor{bluebot}{rgb}{.56862745098039215686,.66274509803921568627,.77254901960784313725}
\definecolor{blueline}{rgb}{.41176470588235294117,.51372549019607843137,.63137254901960784313}
\definecolor{redtop}{rgb}{1,.6,.6}
\definecolor{redbot}{rgb}{1,.276,.276}
\definecolor{redline}{rgb}{.8,0,0}
\definecolor{blueECT}{rgb}{0.8705882353,0.9098039216,0.9490196078}
\definecolor{yellowECT}{rgb}{1,1,0.8431}
\definecolor{hidecolour}{gray}{0.68}

\tikzstyle{bcld} = [stcloud, top color=bluetop, bottom color=bluebot, 
                    blueline,drop shadow={shadow xshift=1pt, shadow
                    yshift=-2pt, opacity=.5, fill=black!30, path
                    fading={circle with fuzzy edge 15 percent},
                    shadow scale=1},rounded corners=0]
\tikzstyle{rcld} = [stcloud, top color=redtop, bottom color=redbot, 
                    redline,drop shadow={shadow xshift=1pt, shadow
                    yshift=-2pt, opacity=.5, fill=black!30, path
                    fading={circle with fuzzy edge 15 percent},
                    shadow scale=1},rounded corners=0]
\tikzstyle{bcld2} = [stcloud, top color=bluetop, bottom color=bluebot, 
                    blueline,rounded corners=0]
\tikzstyle{rcld2} = [stcloud, top color=redtop, bottom color=redbot, 
                    redline,rounded corners=0]
                    
\tikzstyle{region}=[#1,line width=7mm,line cap=round,line join=round]                   
\tikzstyle{animflow}=[blue,draw opacity=0.2,line width=2.7mm,line cap=round,line join=round]                   
\tikzstyle{animnf}=[postaction=decorate,line width=0,draw opacity=0,
                  decoration={markings,mark=at position #1 with 
                  {\node[regular polygon,regular polygon sides=3,rotate=30,draw=red,draw opacity=0.5,line width=1.5pt,line cap=round,line join=round, rounded corners=0.5pt,
                  transform shape,inner sep=1.3pt]{};}}]
\tikzstyle{animfflow}=[blue!20,line width=2.7mm,line cap=round,line join=round]                   

\tikzstyle{actorboxx}=[copy shadow={left color=blue!40},
  left color=blue!40,draw=black,thick]
\tikzstyle{actorbox}=[rectangle, rounded corners=5pt,draw=black!36, fill=blueECT,thick]
\tikzstyle{actorboxtwo}=[rectangle, rounded corners=5pt,draw=black!36, 
  top color=bluetop, bottom color=bluebot!60,thick]
\tikzstyle{yellowbox}=[rectangle, rounded corners=4pt,draw=black!36, fill=yellowECT,thick]
\tikzstyle{whitebox}=[rectangle, rounded corners=0,draw=black!36, fill=white,inner sep=1.3mm,thick]

\newcommand{\mysep}{\,\raisebox{-0.9mm}{\rule{.4pt}{3.5mm}}\,}

\newcommand{\lossy}{Lossy\-Sync\xspace}
\newcommand{\fifo}{FIFO\ensuremath{_1}\xspace}
\newcommand{\sync}{Sync\xspace}
\newcommand{\sdrain}{Sync\-Drain\xspace}

\newcommand{\PS}[1]{\2^{#1}}
\newcommand{\MS}{\mathscr{M}}

\newcommand{\reo}{\ensuremath{\mathsf{Reo}}\xspace}

\newcommand{\dreams}{\ensuremath{\mathsf{Dreams}}\xspace}

\newcommand{\CA}{\textsf{CA}\xspace}

\newcommand{\ie}{i.e.\xspace}

\newcommand{\as}{{\sf AS}\xspace}

\newcommand{\lb}{{\sf L}\xspace}

\newcommand{\bd}{\ba}

\newcommand{\ba}{{\sf BA}\xspace}

\newcommand{\IP}{\ensuremath{\mathit{IP}}\xspace}
\newcommand{\OP}{\ensuremath{\mathit{OP}}\xspace}
\newcommand{\cas}{{\sf CAS}\xspace}

\newcommand{\rell}[2][]{\ell^{(#2)}#1}

\newcommand{\seqc}{\seqc{x}}

\newcommand{\lindaor}{\mathop{\raisebox{-.5mm}{\resizebox{1mm}{2.5mm}{$\boldsymbol{\Box}$}}}}
\newcommand{\Tuple}{\ensuremath{\mathit{Tuple}}\xspace}
\newcommand{\linda}{\ensuremath{\mathsf{Linda}}\xspace}

\newcommand{\dvar}[1]{\ensuremath{\mathscr{#1}}}
\newcommand{\wh}[1]{\ensuremath{\widehat{#1}}\xspace}
\newcommand{\mi}[1]{\mathit{#1}}
\newcommand{\excl}[1]{\mathit{cp}(#1)}
\newcommand{\exclnm}{\ensuremath{\mathit{cp}}\xspace}
\newcommand{\Port}{\X}
\newcommand{\X}{\ensuremath{\mathbb P}\xspace}
\newcommand{\Data}{\ensuremath{\mathbb D}\xspace}
\newcommand{\data}{\ensuremath{\mathit{data}}\xspace}

\newcommand{\Obs}{\dvar{C}\xspace}
\newcommand{\Act}{\ensuremath{\mathit{Act}}\xspace}
\newcommand{\tAct}{\ensuremath{\tau{\mkern-2mu \mathit{Act}}}\xspace}
\newcommand{\DAct}{\ensuremath{\overline{\mathit{Act}}}\xspace}
\newcommand{\Names}{{\ensuremath{\mathscr{N}}}\xspace}

\newcommand{\encd}[1]{\left\llbracket #1\right\rrbracket}

\newcommand{\tpl}[1]{\ensuremath{\left\langle #1\right\rangle}}

\newcommand{\ltpl}{\langle}
\newcommand{\rtpl}{\rangle}
\newcommand{\bigbar}{\ensuremath{~~\big|~~}}
\newcommand{\pmap}{\rightharpoonup}

\newcommand{\scomp}[2]{#1\otimes#2}

\newcommand{\goesby}[1]{\xrightarrow{#1}}

\newcommand{\compat}{\mathop{\raisebox{1mm}{$\frown$}}}

\newcommand{\set}[1]{\ensuremath{\left\{#1\right\}}}

\newcommand{\sset}[1]{\ensuremath{\{#1\}}}

\newcommand{\mset}[1]{\ensuremath{\left\{\!\left|#1\right|\!\right\}}}

\newcommand{\tuple}[1]{\ensuremath{\left\langle #1\right\rangle}}

\newcommand{\HIDE}[1]{}

\newcommand{\wrap}[1]{\begin{tabular}{@{}c@{}}#1\end{tabular}}
\newcommand{\mwrap}[1]{\begin{array}{@{}c@{}}#1\end{array}}

\newcommand{\true}{\mathtt{tt}}

\newcommand{\ttt}{\ensuremath{\mathit{true}}\xspace}
\newcommand{\fff}{\ensuremath{\mathit{false}}\xspace}

\newcommand{\2}{\ensuremath{\mathbf{2}}}

\newenvironment{proof}{\noindent \emph{Proof.\ }}{\hfill$\Box$}

\title{Decoupled execution of synchronous coordination models via behavioural automata
}
\author{
Jos\'e Proen\c{c}a ~~~~~~~~ Dave Clarke
\institute{IBBT-DistriNet, KUL,\\ Leuven, Belgium}
\email{\{jose.proenca,dave.clarke\}@cs.kuleuven.be}
\and
Erik de Vink
\institute{TUE, Eindhoven,\\ The Netherlands}
\email{evink@win.tue.nl}
\and
Farhad Arbab
\institute{CWI, Amsterdam,\\ The Netherlands}
\email{farhad.arbab@cwi.nl}
}

\def\titlerunning{Decoupled execution of synchronous coordination models}
\def\authorrunning{J. Proen\c{c}a, D. Clarke, E. de Vink \& F. Arbab}

\begin{document}
\maketitle

\begin{abstract}
	Synchronous coordination systems allow the exchange of data by logically indivisible
	actions involving all coordinated entities.
  This paper introduces behavioural automata, a logically synchronous coordination model
  based on the Reo coordination language, which focuses on relevant aspects for the concurrent
  evolution of these systems. 
  We show how our automata model encodes the \reo and Linda coordination models and
  how it introduces an explicit predicate that captures the concurrent evolution,
  distinguishing local from global actions, and lifting the need of most synchronous models to involve all entities at each
  coordination step, paving the way to more scalable implementations.
  
\end{abstract}


\section{Introduction}
\label{sec:introduction}

Synchronous constructs in languages such as \reo~\cite{reo} and Esterel~\cite{esterel} are useful for programming reactive systems, though
in general 
their realisations
for coordinating distributed systems
become problematic.
For example, 
it is not clear how to efficiently implement the high degrees of synchronisation expressed by \reo in a distributed context. 
To remedy this situation, the GALS (globally
asynchronous, locally synchronous) model~\cite{gals:phd,gals:verification}
has been adopted, whereby local computation is synchronous and communication
between different machines is asynchronous.

Our work contributes to the field of coordination, in particular to the \reo coordination language, by
incorporating the same ideas behind GALS in our approach to execute
synchronisation models.
More specifically, we introduce \emph{behavioural automata} to model synchronous coordination, inspired in \reo~\cite{reo:ca}.
Each step taken by an automata corresponds to a round of ``synchronous" actions performed by the coordination layer, where data flow atomically through a set of points of the coordinated system.
The main motivation behind behavioural automata is to describe the
synchronous semantics underlying \dreams~\cite{proenca:phd}, a prototype distributed framework
briefly discussed in \myref{sec}{dreams} that stands out by the decoupled execution of
\reo-like coordination models in a concurrent setting.
\dreams improves the performance and scalability of previous attempts to implement similar coordination models.
Our automata model
captures exactly the features implemented by \dreams.
\JP{this paragraph may need some editing.}

Behavioural automata
assume certain properties over their labels, such as the existence of a
composition operator, and use a predicate associated to each of its states that is
needed to guide the composition of automata. Different choices for the
composition operator of labels and the predicates yield different
coordination semantics. We instantiate our automata with the semantics for \reo
and Linda coordination models, but other semantic models can also be captured
by our automata~\cite{proenca:phd}. We do not instantiate behavioural automata with Esterel as the propagation of synchrony in this language differs from our dataflow-driven approach~\cite{reo:interactingcomp}.

Summarising, the main contributions of this paper are:
\JP{mention dreams here?}
\begin{itemize}
  \item   a \emph{unified} automata model that captures dataflow-oriented
  synchronous coordination models;
  \item   the introduction of \emph{concurrency predicates}, increasing the expressiveness of the model when dealing with composed automata; and
  \item   the \emph{decoupling of execution} of a distributed implementation
  based on our automata model, by avoiding unnecessary synchronisation of actions whenever possible.
\end{itemize}
Each behavioural automaton has a concurrenty predicate that indicates, for each state, which labels of other automata require synchronisation. When composing two automata, labels must be either composed in
a pairwise fashion, or they can be performed independently when the concurrency predicate does not require synchronisation.
We exploit how to use concurrency predicates to distinguish transitions of
a composed automaton that originate from all intermediate automata, or from
only a subset of them. We also illustrate how to obtain more complex notions of coordination by increasing the complexity of concurrency predicates.

This paper is organised as follows. We introduce behavioural automata in \myref{sec}{stepwisemodel}. We then encode \reo as behavioural automata in \myref{sec}{reo} and Linda as behavioural automata in \myref{sec}{linda}. In \myref{sec}{cp} we motivate the need for concurrency predicates, both from a theoretical and practical perspectives. We conclude in \myref{sec}{conclusions}.


\section{A stepwise coordination model}
\label{sec:stepwisemodel}

In this section
we present an automata model, dubbed \emph{behavioural automata}.
This model represents our view of a dataflow-driven coordination
system, following the categorisation of Arbab~\cite{reo:interactingcomp}. Each
transition in an automaton represents the \emph{atomic} execution 
of a number of actions by
the coordination system. We describe the behaviour of a system by the \emph{composition} of the behaviour of its sub-systems running
concurrently, each with its own automaton. Furthermore, we allow
the \emph{data values} exchanged over the coordination layer to
influence the choice of how components communicate with each other as well.
We borrow ideas from the Tile model~\cite{tilemodel,reo:tiles},
distinguishing evolution in time (execution of the coordination system) and
evolution in space (composition of coordination systems).
Behavioural automata can be built by \emph{composing} more primitive
behavioural automata, and each transition of an automaton denotes a round of
the coordination process, where data flow \emph{atomically} through zero or
more ports of the system.

We use behavioural automata to give semantics to \reo, based 
on the constraint
automata model~\cite{reo:ca}, and 
to (distributed) Linda~\cite{linda}. Each label of an automaton describes which ports should have dataflow,
and what data should be flowing in each port. We write $\Port$ to denote a
global set of ports, $\lb[P]$ to denote the set of all labels over the ports
$P\subseteq\Port$, and \Data to denote a
global set of data values. We associate a predicate over labels to each
state $q$ of an automaton, referred to as $\Obs(q)$. These predicates are used to guide the composition of behavioural automata.

\begin{definition}[Behavioural automata]
\label{def:ba}
\index{behavioural automaton}
A \emph{behavioural automaton} of a system over a set of ports
$P\subseteq\Port$ is a labelled transition system $\tpl{Q,\lb[P],\to,\Obs}$,
where $\lb[P]$ is the set of labels over $P$, ${\to} \subseteq Q \times \lb[P]
\times Q$ is the transition relation, and $\Obs: Q \to \PS{\lb[P]}$ is a predicate over states and labels, called \emph{concurrency predicate}, regarded as a function that maps states to sets of labels. 
\end{definition}

The key ingredients of behavioural automata are \emph{atomic steps} and
\emph{concurrency predicates}. Each label of a behavioural automaton has an
associated atomic step, which captures aspects such as the ports that have
flow and the data flowing through them, and concurrency predicate describe, for each state,
which labels from other automata running concurrently require synchronisation.

\begin{figure} 
  \centering
  \begin{tabular}{@{}c@{~~~~~~~~~~~~~~~}c@{}c@{}}
    \wrap{\begin{tikzpicture}%
      [node distance=18mm,
              >=triangle 45,thick,bend angle=15,auto,
       smaller/.style={minimum width=15mm},
       further/.style={node distance=23mm},
       closer/.style={node distance=15mm}
      ]    
      \node[main,smaller] (w1) {$W_1$};
      \node[main,closer,below of=w1,smaller] (w2) {$W_2$};
      \draw[-,color=white] (w1) to node (m) {} (w2);
      \node[mycloud,inner sep=3mm,right of=m,black] (ac) {$AC$};
      \node[main,further,right of=ac,smaller] (r) {$R$};
      
      \draw[>-,bend left] (w1.east) to node {$a$} (ac);
      \draw[>-,bend right,swap] (w2.east) to node {$b$} (ac);
      \draw[->] (ac) to node[pos=.4,inner sep=2mm] {$c$} (r);

    \end{tikzpicture}}
    &
    \wrap{\begin{tikzpicture}%
      [shorten >=1pt,>=stealth',node distance=25mm,auto,initial text=,
       bend angle=20,
       eliptic/.style={rectangle,rounded corners=3.5mm,draw=black,minimum
                       height=8mm},
       closer/.style={node distance=10.9mm}]
      \node[eliptic,circle] (e) {$q_0$};
      \node[eliptic,right of=e]  (f) {$q_1(v)$};
      \node[closer,above of=e] (w0) {$\Obs(q_0)$};
      \node[closer,above of=f] (w1) {$\Obs(q_1(v))$};
      \path[->] (e) edge [bend left] node {$s_1(v,w)$}
        (f)
        (f) edge [bend left] node {$s_2(v)$} (e);
      \draw[->,observ] (e) to (w0);
      \draw[->,observ] (f) to (w1);
    \end{tikzpicture}
    }
    &
    \begin{tabular}{r@{~}l}
    $s_1(v,w) = $& read $w$ from $a$,\\
                 & read $v$ from $b$, and\\
                 & write $w$ to $c$\\
    $s_2(v) = $  & write $v$ to $c$
    \end{tabular}
  \end{tabular}
  \caption{Alternating coordinator (left), and its behavioural automaton (right).}
  \label{fig:alternating}
\end{figure}
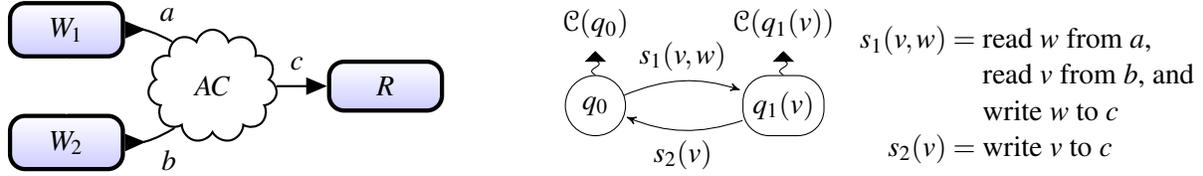

\begin{example}[Alternating coordinator]
\label{ex:ex-ac}

We present the \emph{alternating coordinator} (AC) in \autoref{fig:alternating}.
It receives data from two
data writers $W_1$ and~$W_2$, and sends data to a reader $R$. The
components $W_1$, $W_2$ and~$R$ are connected, respectively, to the ports $a$,
$b$ and $c$ of the alternating coordinator. The alternating coordinator describes
how data can flow between the components, and coordination is specified by the
behavioural automaton depicted on the right side of \autoref{fig:alternating}. Each
transition of this automaton represents a possible step in
time of the coordinator $AC$, describing how the ports $a$, $b$, and $c$ can
have dataflow.
Initially, the coordinator is in state $q_0$, where 
the only possible action is
reading a value $w$ from $W_1$ through $a$ and sending it
to the reader $R$ through $c$, while reading and buffering a value $v$ sent by $W_2$ through $b$. Note
that if only one of the writers can produce data, the step cannot be taken, and
the system cannot evolve. In the next state, $q_1$, 
the only possible step is to send the value $v$ to the reader $R$, returning to state~$q_0$.
The
arrows between states
represent the transition relation $\to$. In both states there is
the possibility of allowing the concurrent execution of other automata, provided that this
execution does not interfere with the current behaviour. The conditions
of when other automata can execute concurrently 
are captured by the concurrency predicate~$\Obs$, 
depicted
by squiggly arrows
(\raisebox{0.5mm}{\tikz \draw[->,observ] (0,0) -- (0.9,0);}) from 
each state.
\end{example}

\subsection{Labels, atomic steps and concurrent predicates}
\label{sec:atomicstep}

Labels over a set of ports $P$ are elements from a set $\lb[P]$ with some properties required for composition, which we will introduce later.
Furthermore, a label $\ell \in \lb[P]$ can be restricted to a smaller set of ports $P'\subseteq P$, written $\rell{P'}$.
We require each label $\ell \in \lb[\Port]$ to have an associated description of where and which data flow in the connector, written as $\alpha(\ell)$, and captured by the notion of \emph{atomic step}.

\begin{definition}[Atomic step]
\label{def:as}
An \emph{atomic step} over the alphabet $P\subseteq \X$ is a tuple
$\tpl{P,F,\IP,\OP,\data}$ where:
\\$
~~~~~~
\begin{array}{r@{~}l@{~~~~~~~~~~}r@{~}l@{~~~~~~~~~~}r@{~}l@{~~~~~~~~~~}r@{~}l@{~~~~~~~~~~}r@{~}l}
F &\subseteq P
&
IP &\subseteq F
&
OP &\subseteq F
&
IP \cap \OP &= \emptyset
&
\text{and~~~~~~~}data :& (\IP\cup\OP)\to \Data.
\end{array}$
\end{definition}

\noindent
We write $\as[P]$ to denote the set of all atomic steps over the ports in $P$.
$P$ is a set of ports in the scope of the atomic step. The flow set $F$ is the set of ports that \emph{synchronise}, \ie, that have
data flowing in the same atomic step. The sets \textit{IP} and \textit{OP}
represent the input and output ports of the atomic step that have dataflow, and
whose values are considered to be relevant when performing a step.
Ports in $F$ but not in \IP or \OP are ports with dataflow, but whose data
values are not relevant, that is, they are used only for imposing
synchronisation constraints. The data values that flow through the relevant
ports are given by the data function \data. We distinguish \IP and \OP to
capture data dependencies.

Concurrency predicates are used to compose behavioural automata. When
composing two automata $a_1$ and $a_2$, if $a_1$ has ports $P_1$, has the concurrency predicate $\Obs_1$, and is in state $q_1$,
then
$\rell[_2]{P_1} \in \Obs_1(q_1)$ means that
$a_2$ can perform $\ell_2$ only when composed with a transition from $a_1$,
otherwise
$a_2$ can perform
$\ell_2$ without requiring $a_1$ to perform a transition.%
\footnote{We present a variation of the original definition of concurrency predicates~\cite{proenca:phd} to make the decision of belonging to a concurrent predicate more local.}
When clear from context, we omit the restriction and write $\ell_2 \in \Obs_1(q_1)$ instead of $\rell[_2]{P_1} \in \Obs_1(q_1)$.
We give a possible definition for concurrency predicates based solely on the set of
known ports.\footnote{Other semantic models may require more complex concurrency predicates. For example, the concurrency
predicates for the \reo automata model~\cite{reo:ra} depend on the current state
(Section~3.6.2 of~\cite{proenca:phd}).
} Given a connector with known ports $P_0$, the concurrency
predicate of every state is given by the predicate
\vspace{-2mm}
\begin{align}
\excl{P_0} ~=~ \set{\ell ~|~ \alpha(\ell) = \tpl{P,F,\IP,\OP,\data}, P_0 \cap F \neq \emptyset}.
\label{eq:excl}
\end{align}

\begin{example}
We define the atomic steps and concurrency predicates from \myref{ex}{ex-ac} as follows.
\vspace{-2mm}
\[
\begin{array}{@{\alpha}l@{~=~}l@{, }r@{, }r@{, }r@{, }l@{~~~~~~~~~~~}l@{~=~}l}
      (s_1(v,w))
      & \langle P & abc & ab & c & \set{a,b,c\mapsto w,v,w}\rangle 
      &
      \Obs(q_1(v)) & \excl{P}
      \\
      (s_2(v)) &
      \langle P & c & \emptyset & c & \set{c\mapsto v}\rangle 
      &
      \Obs(q_0) & \excl{P} 
\end{array}\]
\vspace{-5mm}

For simplicity, we write $a_1\ldots a_n$ instead of $\set{a_1,\ldots,a_n}$ 
when 
the intended notion of set
is clear from the context.
The alphabet $P$ is $\set{a,b,c}$, and the
concurrency predicates allow only steps where none of the known ports has flow.
\end{example}

\subsection{Composition of behavioural automata}
\label{sec:bacomp}

To compose behavioural automata we require labels to be elements of a partial monoid
$\tpl{\lb,\scomp{}{}}$, that is,
(1) there must be a commutative operator $\scomp{}{} : \lb^2 \pmap \lb$ for labels,
and (2) the composition of two labels can
be undefined, meaning that they are incompatible. For technical convenience,
we require $\scomp{}{}$ to be associative and to have an identity element.
The atomic step $\tpl{P,F,\IP,\OP,\data}$ of a composed label $\scomp{\ell_1}{\ell_2}$ must obey
the following conditions,
where, for every label $\ell_1$ or $\ell_2$, $\alpha(\ell_i) = \tpl{P_i,F_i,\IP_i,\OP_i,\data_i}$.
\vspace{-2mm}
\[\begin{array}{r@{~}l@{\hspace{20mm}}r@{~}l@{\hspace{20mm}}r@{~}l}
P &\subseteq P_1 \cup P_2
&
\IP &\subseteq (\IP_1 \cup \IP_2) \backslash (\OP_1 \cup \OP_2)
&
\data_1 \compat& \data_2
\\
F &\subseteq F_1 \cup F_2
&
\OP &\subseteq \OP_1 \cup \OP_2
&
\data = \data_1 &\cup~ \data_2
\end{array}\]
\vspace*{-5mm}

\reviewin{3: It is not clear to me why the connected input ports are removed
from a composition of two automata ("since the dependencies have been met")
while the output ports persist.}
The atomic step of a label
$\ell$ is represented by $\alpha(\ell)$.
The notation $m_1 \compat m_2$ represents that the values of the common domain
of mappings $m_1$ and $m_2$ match. The requirements on the sets \IP and \OP reflect that when composing
two atomic steps, the input ports that have an associated output port are no
longer treated as input ports (since the dependencies have been met), and the
output ports are combined.
The intuition behind the removal of input ports that match an output port is
the preservation of the semantics of \reo: multiple connections to an output
port replicate data, but multiple connections to input data require the
merging of data from a single source.
\JP{new text}

We now describe the composition of behavioural automata based on the
operator $\scomp{}{}$ and on concurrency predicates.
This composition
mimics the composition of existing \reo models~\cite{reo:ca,reo:cc,reo:ra}.

\index{composition of!behavioural automata ($\bowtie$)}
\begin{definition}[Product of behavioural automata] \label{def:bacomp}
The product
of two behavioural automata $b_1=\langle Q_1,$ $\lb[P_1],\to_1,\Obs_1\rangle$ and
$b_2=\ltpl Q_2,\lb[P_2], \to_2,\Obs_2 \rtpl$, denoted by $b_1 \bowtie b_2$, is the behavioural automaton
$\ltpl Q_1\times Q_2, \lb[P_1 \cup
P_2],\to,\Obs \rtpl$, where
$\to$ and $\Obs$ are  defined as follows:
\vspace{-2mm}
\begin{align}
\hspace*{-1cm}{\to} & ~=~  
\sset{\tpl{(p,q),\ell,(p',q')} ~|~ p \xrightarrow{\ell_1}_1 p', q \xrightarrow{\ell_2}_2
  q', \ell = \scomp{\ell_1}{\ell_2} , \ell\neq \bot}
\label{eq:comp1}
~\cup \\
&
\sset{\tpl{(p,q),\ell,(p',q)} ~|~ p \xrightarrow{\ell}_1 p', \rell{P_2}\notin \Obs_2(q)}
~\cup~
\sset{\tpl{(p,q),\ell,(p,q')} ~|~ q \xrightarrow{\ell}_2 q', \rell{P_1}\notin \Obs_1(p)}
\label{eq:comp3}
\\
\hspace*{-1cm}
\rule{0pt}{14pt}
\Obs(p,q) &~=~
  \Obs_1(p) \cup \Obs_2(q) \textrm{~~~for~~~} p \in Q_1, q \in Q_2.
\label{eq:nfcomp}
\end{align}
\end{definition}

Case
(\ref{eq:comp3}) covers the situation where one of
the behavioural automata performs a step admitted by the concurrency predicate
of the other, and case~(\ref{eq:nfcomp}) defines the composition of two
concurrency predicates. 

In practice, our framework based on behavioural automata, briefly described in \myref{sec}{dreams}, uses a symbolic
representation for data values assuming that variables can be instantiated
after selecting the transition. This suggests the use of a late-semantics for
data-dependencies.
Our approach to compose labels resembles Milner's
synchronous product in SCCS~\cite{sccs}, with the main difference that the
product of behavioural automata do not require the all labels to be
synchronised. The product of labels from two behavioural automata can be
undefined, and labels can avoid synchronisation when the concurrency predicate
holds.
\JP{new text.}

\reviewin{1: How is the definition of the product of behavioral automata related to synchronous product previously defined in the literature?}
\JPin{We say, for the CA, that we expect to be the same, but we did not formally prove it. Dave: mention SCSS - it involves everyone to engage in a lock step, while our approach requires only a subset of the system (and SCCS is a framework...).\\
We now mention SCCS in a sentence.}

\subsection{Example: lossy alternator}
\label{sec:lossyalternator}

Recall the behavioural automaton $\mi{AC}$ of the
alternating coordinator, illustrated in \myref{fig}{alternating}.
Data is received
always via ports $a$ and $b$ simultaneously, and sent via port $c$,
alternating the values received from $a$ and $b$. We now imagine the following
scenario: the data on $a$ becomes available always at a much faster rate than
data on $b$. To adapt our alternating coordinator to this new scenario, we introduce a lossy-FIFO
connector $\mi{LF}$~\cite{reo} and compose it with the alternating coordinator, yielding 
$\mi{LF}\bowtie \mi{AC}$.

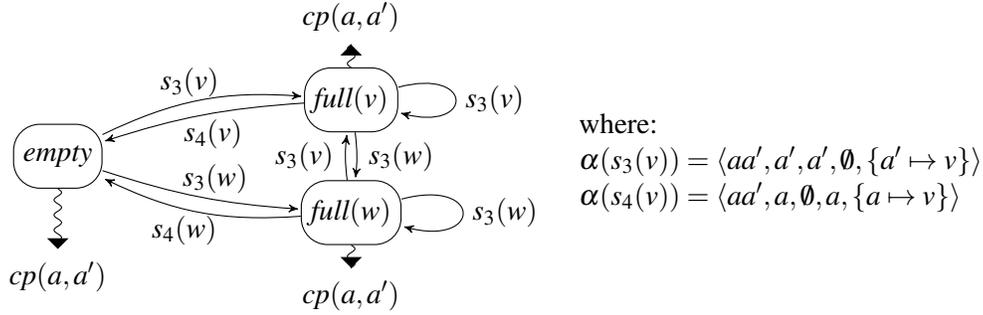
\begin{figure}
  \centering
  \begin{tabular}{@{}cc@{}}
    \wrap{\begin{tikzpicture}%
      [shorten >=1pt,>=stealth',node distance=39mm,auto,initial text=,
       bend angle=7,
       obsdist/.style={node distance=11mm},
       closer/.style={node distance=7.5mm},
       slabel/.style={inner sep=0mm, pos=.4},
       outa/.style={bend angle=15},
       ina/.style={bend angle=-7},
       eliptic/.style={rectangle,rounded corners=3.5mm,draw=black,minimum height=8.5mm}]
      \node[eliptic] (e) {$\mathit{empty}$};
      \node[right of=e] (m) {};
      \node[eliptic,closer,above of=m] (fv) {$\mathit{full}(v)$};
      \node[eliptic,closer,below of=m] (fw) {$\mathit{full}(w)$};
      \node[obsdist,above of=fv] (fvobs) {$\excl{a,a'}$};
      \node[obsdist,below of=fw] (fwobs) {$\excl{a,a'}$};
      \node[below of=e,node distance=16mm] (eobs) {$\excl{a,a'}$};
      \path[->] (e)  edge [slabel,outa,bend left,pos=.6] node {$s_3(v)$} (fv)
                (e)  edge [slabel,ina, bend left,pos=.4] node {$s_3(w)$} (fw)
                (fv) edge [slabel,ina, bend left,pos=.6] node {$s_4(v)$} (e)
                (fw) edge [slabel,outa,bend left,pos=.4] node {$s_4(w)$} (e)
                (fv) edge [inner sep=1.5mm,bend left,pos=.5] node {$s_3(w)$} (fw)
                (fw) edge [inner sep=1.5mm,bend left,pos=.5] node {$s_3(v)$} (fv)
                (e)  edge [observ] (eobs)
                (fv)  edge [loop right] node {$s_3(v)$} ()
                (fw)  edge [loop right] node {$s_3(w)$} ()
                (fv) edge [observ] (fvobs)
                (fw) edge [observ] (fwobs);
    \end{tikzpicture}
    }
    &
    \begin{tabular}{@{}l@{}}
      where:\\
      $\alpha(s_3(v)) = \tpl{aa',a',a',\emptyset,\set{a'\mapsto v}}$\\
      $\alpha(s_4(v)) = \tpl{aa',a,\emptyset,a,\set{a\mapsto v}}$\\
    \end{tabular}
  \end{tabular}
  \caption{Behavioural automaton of the lossy-FIFO connector.}
  \label{fig:lossyfifo}
  \index{connector!lossy-FIFO}
\end{figure}

Recall the definition of $\exclnm:\Port \to \lb[\Port]$ given by \myref{eq}{excl}. The behavioural automaton for the lossy-FIFO connector
is depicted in \myref{fig}{lossyfifo}, and its atomic steps range over the
ports $\set{a,a'}$, where $a'$ is an input port and $a$ is an output
port.
We depict the interface of both of these connectors on left hand side of
\myref{fig}{newalternating}. After
combining the behavioural automata of the two connectors, they
become connected via their shared port $a$. The new variation of the alternating
coordinator can then be connected to data producers and consumers by using the ports
$a'$, $b$ and $c$, as depicted at the right hand side of \myref{fig}{newalternating}.

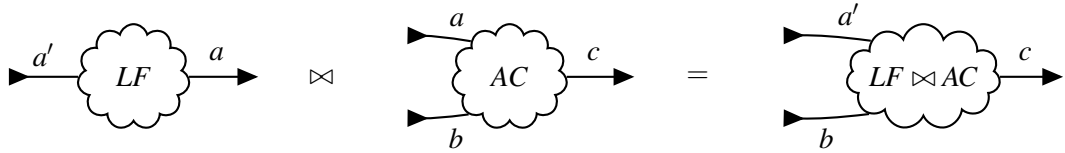
\begin{figure}
  \centering
  \begin{tikzpicture}%
      [node distance=28mm,
              >=triangle 45,thick,bend angle=4,auto,
       further/.style={node distance=18mm},
       closer/.style={node distance=5.5mm}
      ]    
      \node[] (w1) {};
      \node[mycloud,inner sep=0mm,minimum height=14mm,
        right of=w1,black,further] (lf) {$~~~\mi{LF}~~~$};
      \node[further,right of=lf] (r2) {};
      \draw[>-] (w1) to node {$a'$} (lf);
      \draw[->,inner sep=2mm,pos=.4] (lf) to node {$a$} (r2);

      \node[right of=lf, node distance=25mm] (join) {$\bowtie$};

      \node[mycloud,inner sep=0mm,minimum height=14mm,
        right of=join,black,node distance=25mm] (ac) {$~~~\mi{AC}~~~$};
      \node[left of=ac,node distance=15mm] (m) {};
      \node[closer, above of=m] (w1) {};
      \node[closer,below of=m] (w2) {};
      \node[further,right of=ac] (r) {};
      \draw[>-,bend left] (w1.east) to node {$a$} (ac.north west);
      \draw[>-,bend right,swap] (w2.east) to node {$b$} (ac.south west);
      \draw[->,inner sep=2mm,pos=.4] (ac) to node {$c$} (r);

      \node[right of=ac, node distance=25mm] (eq) {$=$};

      \node[mycloud,inner sep=0mm,minimum height=14mm,
        right of=eq,black,node distance=30mm] (acx) {$LF \bowtie AC$};
      \coordinate[node distance=20mm,left of=acx] (mx) {};
      \node[node distance=5.5mm,above of=mx] (w1x) {};
      \node[node distance=5.5mm,below of=mx] (w2x) {};
      \node[node distance=20mm,right of=acx] (r) {};
      
      \draw[>-,bend left] (w1x.east) to node {$a'$} (acx.north west);
      \draw[>-,bend right,swap] (w2x.east) to node {$b$} (acx.south west);
      \draw[->,inner sep=2mm,pos=.4] (acx) to node {$c$} (r);
  \end{tikzpicture}
  \caption{
  The sink and source ports of $\mi{LF}$, $\mi{AC}$, and their composition.}
  \label{fig:newalternating}
\end{figure}
\index{connector!alternating}

Intuitively, the lossy-FIFO connector receives data $a'$
and buffers its value before sending it through~$a$. When the buffer is
full data received from $a'$ replaces the content of the buffer.
\JPin{Farhad:
What you describe here is the behavior of the shift-lossy-FIFO1 connector.  Everywhere else in this paper, you actually use the overflow-lossy-FIFO1 connector.  Strictly speaking, this is fine, because these are different examples.  Nevertheless, I think this confuses the reader.  Either say something about this (perhaps in a footnote here or in the later examples) or change this statement to describe the behavior of an overflow-lossy-FIFO1 as well.}
\JPin{Addressed later.}
The connector resulting from the
composition $\mi{LF} \bowtie \mi{AC}$ is formalised in \myref{tab}{ac+lf} and
in \myref{fig}{ac+lf}.
The flow sets of the labels $s_1(v,w)$, $s_2(v)$, $s_3(v)$
and $s_4(v)$ are, respectively, $abc$, $c$, $a'$, and $a'a$, and the set of known
ports is $P = \set{a',a,b,c}$.
Let $\Obs_{LF}$ and $\Obs_{AC}$ be the concurrency predicates of $LF$ and $AC$. The concurrency predicate $\Obs_{LF\bowtie AC}$
for $LF\bowtie AC$ results from the union of the predicates of the
states of each behavioural automaton, and corresponds precisely to the
concurrency predicate that maps each state to $\excl{a',a,b,c}$.
The name of each state in $LF\bowtie AC$ is obtained by pairing names of a state from $LF$ and a state from $AC$.
Some states and transitions are coloured in grey with their labels omitted to avoid cluttering the diagram.

\reviewin{2: "the set of known ports is always {a', a, b, c}" is very
unintuitive...I would expect the known ports of LF to be {a', a} and the known
ports of AC to be {a, b, c}. There's something weird going on when the set of
known ports expands when you put two components together. That's one reason I
think this way of specifying concurrency predicates is backwards.}
\JPin{the set of ports is always constant in our example. We can also hide
intermediate ports, as it is done for reo automata. And what is so weird about
expanding the set of know ports when composing 2 automata?\\
I dropped "always", and now concurrency predicates are inverted and localised.}

From the diagram it is clear that some transitions originate only from the
$LF$ or the $AC$ connector, while others result from the composition via
the operator $\scomp{}{}$. The transitions $s_2(v)$ and
$s_3(w)$ can be performed simultaneously or interleaved; simultaneously
because $\scomp{s_2(v)}{s_3(w)}$ is defined, and interleaved because
$\Obs_{\it LF}$ never contains $s_2(v)$
and $\Obs_{\it AC}$ never contains
$s_3(w)$. The possible execution scenarios of these atomic steps follow our
intuition that steps `approved' by concurrency
predicates can be performed independently. The steps $s_1(u,v)$ and
$s_4(w)$ can be taken only when composed.

\def\firsttablerule{\hrule \@height 1pt 
}
\def\thickhrulefill{\leavevmode \leaders \hrule height 1pt\hfill \kern \z@}

\newcommand*{\thickhrulefilll}{%
 \leavevmode 
 \leaders\hrule height 1pt\hfill 
 \kern 0pt\relax 
}

\begin{table}
  \centering
  $\begin{array}{c|c@{\hspace{1mm}}c}
  \mytoprule
  \scomp{}{} & s_1(u,v) & s_2(w)
  \mymidrule
  s_3(y)
    & \bot & \mwrap{\ltpl P,a'c,a',c,~~~~~\\~\set{a',c\mapsto y,w} \rtpl}\\
  s_4(z)
    & \bot \text{ (for $z\neq v$)} & \bot\\
  s_4(v)
    & \mwrap{\ltpl P,abc,ab,c,~~~~~~~\\~~\set{a,b,c\mapsto v,u,v} \rtpl} & \bot%
  \mybottomrule
  \end{array}$
~~~~~~~~~~~~
$\begin{array}{c|cc}
  \mytoprule
  LF & \Obs_{LF}(\mathit{empty}) & \Obs_{LF}(\mathit{full}(v'))
  \mymidrule
  s_1(v,w)
    & \ttt & \ttt\\
  s_2(v)
    & \fff & \fff%
  \mybottomrule
  \multicolumn{3}{c}{~} \\
  \mytoprule
  AC & \Obs_{AC}(q_0) & \Obs_{AC}(q_1(v'))
  \mymidrule
  s_3(v)
    & \fff & \fff\\
  s_4(v)
    & \ttt & \ttt%
  \mybottomrule
  \end{array}$  
  \caption{Atomic steps of the composition of labels from $LF$ and $AC$ (left), and verification of the concurrency predicate for each label (right).}
  \label{tab:ac+lf}
\end{table}

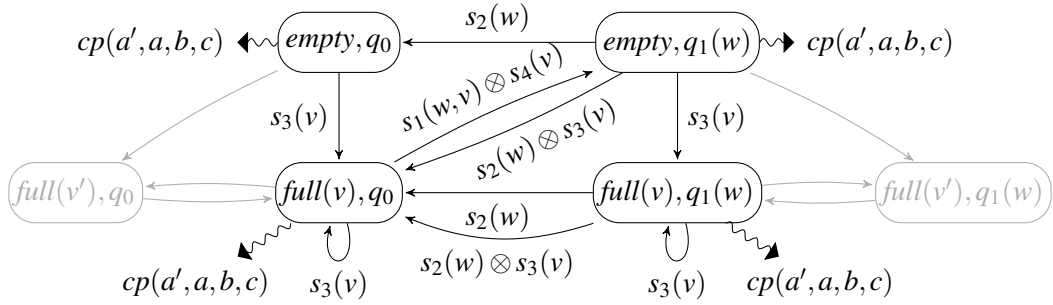
\begin{figure}
  \centering
  \begin{tikzpicture}%
    [shorten >=1pt,>=stealth',node distance=20mm,auto,initial text=,
     bend angle=5,
     lgray/.style={hidecolour},
     further/.style={node distance=45mm},
     closer/.style={node distance=12mm},
     eliptic/.style={rectangle,rounded corners=3.5mm,draw=black,minimum height=8mm,inner sep=1mm}]
    \node[eliptic] (eq0) {$\mathit{empty},q_0$};
    \node[eliptic,further,right of=eq0] (eq1) {$\mathit{empty},q_1(w)$};
    \node[eliptic,below of=eq0] (fq0) {$\mathit{full}(v),q_0$};
    \node[eliptic,below of=eq1] (fq1) {$\mathit{full}(v),q_1(w)$};  
    \node[node distance=25mm,left of=eq0] (eq0obs)
      {$\excl{a',a,b,c}$};
    \node[node distance=27mm,right of=eq1] (eq1obs)
      {$\excl{a',a,b,c}$};
    \node[closer, below of=fq0] (mfq0) {};
    \node[node distance=19mm,left of=mfq0] (fq0obs)
      {$\excl{a',a,b,c}$};
    \node[closer,below of=fq1] (mfq1) {};
    \node[node distance=19mm,right of=mfq1] (fq1obs)
      {$\excl{a',a,b,c}$};
    \node at ($(eq0)!.5!(fq0)$) (mq0) {};
    \node[node distance=35mm,eliptic,lgray,left of=fq0] (fq0w)
      {$\mathit{full}(v'),q_0$};
    \node at ($(eq1)!.5!(fq1)$) (mq1) {};
    \node[node distance=38mm,eliptic,lgray,right of=fq1] (fq1w)
      {$\mathit{full}(v'),q_1(w)$};
    \draw[->,lgray, bend right] (eq0) to (fq0w);
    \draw[->,lgray, bend right] (fq0w) to (fq0);
    \draw[->,lgray, bend right] (fq0) to (fq0w);
    \draw[->,lgray, bend left] (eq1) to (fq1w);
    \draw[->,lgray, bend left] (fq1w) to (fq1);
    \draw[->,lgray, bend left] (fq1) to (fq1w);
    \draw[->,observ] (eq0) to (eq0obs);
    \draw[->,observ] (eq1) to (eq1obs);
    \draw[->,observ] (fq0) to (fq0obs);
    \draw[->,observ] (fq1) to (fq1obs);
    \path[->] (eq0) 
                    edge [swap]       node {$s_3(v)$} (fq0)
              (fq0) 
                    edge [loop below] node {$s_3(v)$} ()
                    edge [bend left,sloped,above]  node 
                      {$\scomp{s_1(w,v)}{s_4(v)}$} (eq1)
              (eq1) 
                    edge [swap]       node {$s_2(w)$} (eq0)
                    edge [bend left,sloped,below]  node [pos=.4]
                      {$\scomp{s_2(w)}{s_3(v)}$} (fq0)
                    edge              node {$s_3(v)$} (fq1)
              (fq1) 
                    edge [loop below]  node {$s_3(v)$} ()
                    edge [] node {$s_2(w)$} (fq0)
                    edge [bend angle=20,bend left]  node 
                      {$\scomp{s_2(w)}{s_3(v)}$} (fq0);
  \end{tikzpicture}
  \caption{Behavioural automaton for the composition of $LF$ and $AC$.}
  \label{fig:ac+lf}
\end{figure}

\subsection{Locality} 
\label{sec:locality and grouping}

\index{locality}
\index{step!local}

We introduce the notion of locality as a property of behavioural automata
that
guarantees the absence of certain labels in the concurrency predicates of \emph{independent} behavioural automata, that is, in automata without shared ports.

\begin{definition}[Locality of behavioural automata]
\label{def:localityba}
A behavioural automaton $b = \tpl{Q,\lb[P],\to,\Obs}$ obeys the \emph{locality property} if, for any port set $P'$ such that $P\cap P'=\emptyset$,
$~~\forall \ell \in \lb[P'] \cdot \forall q \in Q \cdot \rell{P} \notin \Obs(q)$.
\end{definition}

Any two behavioural automata with disjoint port sets that obey the locality
property can therefore evolve concurrently in an interleaved fashion. Let
$b=b_1\bowtie b_2$ be a behavioural automaton
and $\ell$ a label from~$b_1$. We say $\ell$ is a \emph{local step} of $b$ if
$(q_1,q_2) \goesby{\ell} (q'_1,q_2')$ is a transition of $b$ and either $q_1
\goesby{\ell}_1 q'_1$, $q_2 = q_2'$, and
$\ell \in \Obs_2(q_2)$; or $q_2 \goesby{\ell}_2 q'_2$, $q_1=q_1'$, and $\ell \in \Obs_1(q_1)$.
In the behavioural automaton exemplified in
\myref{fig}{ac+lf}, the local steps are exactly the transitions labelled by
the steps $s_2(w)$ and $s_3(v)$.
\begin{proposition} \label{prop:localstep}
Let $b = b_1 \bowtie b_2 \bowtie b_3$ be a
behavioural automaton 
where $b_i = \tpl{Q_i,\lb[P_i],\to_i,\Obs_i}$, for $i\in
1..3$,
and assume the locality property from \myref{def}{localityba} holds for $b_1$, $b_2$ and $b_3$.
Suppose $P_1 \cap P_3 = \emptyset$. Then, for any step $\rell[_1]{P_1} \in \lb[P_1]$ performed by $b_1$ and $q_2 \in Q_2$,
if $\rell[_1]{P_2} \notin \Obs_2(q_2)$  then $\ell_1$ is a local step of $b$.
\end{proposition}

\JPin{From a comment of Dave: is $\bowtie$ associative? Actually, we require that $\scomp{s_1}{s_2} \in \Obs(q_3) \leftrightarrow s_1,s_2 \notin \Obs(q_3)$. Shall we state this somewhere? (no space!)}

\begin{proof}
Observe that $\bowtie$ is associative, up to the state names, because the composition of labels $\scomp{}{}$ is associative.
From $P_1\cap P_3 = \emptyset$, $\ell_1 \in \lb[P_1]$, and from the locality
property in \myref{def}{localityba} we conclude that $\forall q \in Q_3 \cdot
\rell[_1]{P_3} \notin \Obs_3(q)$. Therefore, for any state $q_3 \in Q_3$ and for a state
$q_2 \in Q_2$ such that $\rell[_1]{P_2} \notin \Obs_2(q_2)$, we have that $\rell[_1]{P_2} \notin
\Obs_2(q_2) \cup \Obs_3(q_3)$. We conclude that $\rell[_1]{P_2\cup P_3} \notin \Obs'$, where $\Obs'$
is the concurrency predicate of $b_2 \bowtie b_3$, hence a local step of $b$.
\end{proof}

If the locality property
holds for each behavioural automata $b_i$ in a composed system $b =
b_1\bowtie \cdots \bowtie b_n$, then, using \myref{prop}{localstep}, we can infer wether atomic steps from
$b_i$ are local steps of $b$ based only on the concurrency predicates of its
\emph{neighbour automata}, \ie, the automata that share ports with $b_i$.

\subsection{Concrete behavioural automata}
\label{sec:instantiation}

A behavioural automaton is an abstraction of concrete coordination models that focuses on aspects relevant
to the execution of the coordination model.
As we will argue, \reo and Linda can 
be cast in our framework of behavioural automata.
Therefore, both \reo and Linda coordination models can be seen as specific instances of the stepwise model described above.
For a concrete coordination model to fit into the stepwise model, we need to define:
  (1) labels in the concrete model;
  (2) the encoding $\alpha$ of labels into atomic steps;
  (3) composition of labels; and
  (4) concurrency predicates. 
  \JP{We do not mention restriction...}

We start by encoding the constraint automata semantics of \reo as behavioural automata. Later,
because of its relevance in the coordination community as one of the first
coordination languages, we also encode Linda as a behavioural automaton.
Other coordination models have also been encoded as behavioural automata in Proen\c{c}a's Ph.D. thesis~\cite{proenca:phd}.


\section{Encoding Reo} 
\label{sec:reo}     

\reo~\cite{reo,reo:abt} is presented as a channel-based coordination language
wherein component connectors are compositionally built out of an open
set of \emph{primitive connectors}, also called primitives. Channels
are primitives with two ends.
Existing tools for \reo include an editor, an
animation generator, model checkers, editors of \reo-specific automata, QoS
modelling and analysis tools, and a code generator~\cite{reo:ect,christian:phd}.

The behaviour of each primitive depends upon
its current state.\footnote{Note that most \reo primitives presented here have a single state.} The semantics of a connector is described as a collection of possible steps for each state, and we call the change of state of the connector triggered by one of these steps a \emph{round}.
At each round some of the ends of a connector are synchronised, \ie, only certain combinations of synchronous dataflow through its ends are possible.
Dataflow on a primitive's end occurs
when a single datum is passed through that end. Within any round dataflow may
occur on some number of ends.
Communication with a primitive connector occurs through its ports, called
\emph{ends}. Primitives consume data through their \emph{source ends}, and
produce data through their \emph{sink ends}.
Connectors are formed by plugging the ends
of primitives together in a one-to-one fashion 
to form \emph{nodes}. A node is a logical place consisting of a sink end,
a source end, or both a sink and a source end.\footnote{Generalised nodes with  multiple sink and source
ends can be defined by combining binary mergers and
replicators~\cite{reo:ca,reo:cc}.} 

We now give an informal description of some of the most commonly used \reo
primitives. Note that, for all of these primitives, no dataflow is one of the
behavioural possibilities.
The \sync channel (\chn{sync}) sends data synchronously from its source to its
sink end. The \lossy channel (\chn{lossy}) differs from the \sync channel only
because it can non-deterministically lose data received from its source port.
The \sdrain (\chn{sdrain}) has two source ends, and requires both ends to have
dataflow synchronously, or no dataflow is possible. The \fifo channel (\chn{fifo}) has
two possible states: empty or full. When empty, it can receive a data item from
its source end, changing its state to full. When full, it can only send the
data item received previously, changing its state back to empty. Finally,
a replicator (\chnsmallreplicator) replicates data synchronously to all of its
sink ends, while a merger (\chnsmallmerger) copies data atomically from exactly
one of its sink ends to its source end.

\medskip

\noindent
\begin{minipage}{.75\textwidth}
\begin{example} \label{ex:exrouter}
\index{connector!exclusive router}
The connector on the right
is an exclusive router built by composing
two LossySync channels ($b$-$e$ and $d$-$g$), one SyncDrain ($c$-$f$), one
Merger ($h$-$i$-$f$), and three Replicators ($a$-$b$-$c$-$d$, $e$-$j$-$h$ and $g$-$i$-$k$). 
The constraints of these primitives can be combined to give the following two
behavioural possibilities (plus the no-flow-everywhere possibility):
\begin{itemize}
  \item ends $\set{a,b,c,d,e,f,h,j}$ synchronise and a data item flows from 
  $a$ to $j$,
  \item ends $\set{a,b,c,d,f,g,i,k}$ synchronise and a data item flows from
  $a$ to $k$.
\end{itemize}
\end{example}
\end{minipage}
\tikzstyle{erlength} = [node distance=5.2mm]
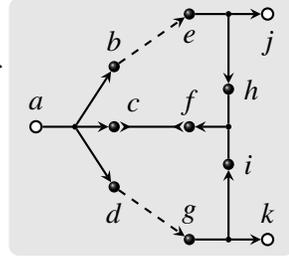
\begin{minipage}{.25\textwidth}
    \reoconnector{
    \node[boundary,label={[name=alb]above:$a$}] (a) {};
    \node[point,erlength,right of=a] (m1) {};
    \node[mixed,erlength,right of=m1,label=60:$c$] (c) {};
    \node[mixed,node distance=8mm,above of=c,label=above:$b$] (b) {};
    \node[mixed,node distance=8mm,below of=c,label=below:$d$] (d) {};
    \node[mixed,node distance=10mm,right of=c,label={[inner sep=2pt]above:$f$}] (f) {};
    \node[mixed,node distance=15mm,above of=f,label=below:$e$] (e) {};
    \node[mixed,node distance=15mm,below of=f,label=above:$g$] (g) {};
    \node[point,erlength,right of=e] (m2) {};
    \node[point,erlength,right of=f] (m3) {};
    \node[point,erlength,right of=g] (m4) {};
    \node[mixed,node distance=5mm,above of=m3,label=right:$h$] (h) {};
    \node[mixed,node distance=5mm,below of=m3,label=right:$i$] (i) {};
    \node[boundary,erlength,right of=m2,label={[name=jlb]below:$j$}] (j) {};
    \node[boundary,erlength,right of=m4,label={[name=klb]above:$k$}] (k) {};
    \draw[sync] (a)--(m1)--(b) {};
    \draw[sync]    (m1)--(c) {}; \draw[sync]    (m1)--(d) {};
    \draw[lossy]   (b)--(e)  {}; \draw[lossy]    (d)--(g) {};
    \draw[channel] (e)--(m2) {}; \draw[channel] (g)--(m4) {};
    \draw[sync]    (m2)--(h) {}; \draw[sync]    (m4)--(i) {};
    \draw[sdrain]   (c)--(f) {}; \draw[sync]    (m3)--(f) {};
    \draw[channel] (h)--(m3) {}; \draw[channel] (i)--(m3) {};
    \draw[sync]    (m2)--(j) {}; \draw[sync]    (m4)--(k) {};
    \reobox{(alb)(jlb)(klb)(j)(k)}
  }
\end{minipage}

\vspace{2.5mm}

The merger makes a non-deterministic choice whenever both behaviours are possible.
Data can never flow from $a$ to both $j$ and $k$, as this is excluded
by the behavioural constraints of the Merger $h$-$i$-$f$.


\subsection{Constraint automata}
\label{sec:ca}

We briefly describe constraint automata~\cite{reo:ca}. Constraint automata use a finite set of port names $\Names =
\set{x_1,\ldots,x_n}$, where $x_i$ is the $i$-th port of a connector.
When clear from the context, we write $xyz$ instead of $\set{x,y,z}$ to
enhance readability.
We write $\wh{x_i}$ to represent the variable that holds the data value
flowing through the port $x_i$, and use $\wh\Names$ to denote the set of data variables $\set{\wh{x}_1,\ldots,\wh{x}_n}$, for
each $x_i\in \Names$. We define $DC_{X}$ for each $X \subseteq \Names$ to be a
set of data constraints over the variables in $\wh{X}$, where the underlying data
domain is a finite set \Data.
Data constraints in $DC_{\Names}$ can be
viewed as a symbolic representation of sets of data-assignments, and are
generated by the following grammar:
\index{constraints!data constraints ($DC_X$)}
\vspace*{-2mm}
\[
g ~~::=~~ \true ~\bigm|~ \wh{x} = d ~\bigm|~ g_1 \lor g_2 ~\bigm|~ \lnot g
\vspace*{-2mm}
\]
where $x \in \Names$ and $d \in \Data$. The other logical connectives can be
encoded as usual. We use the notation $\wh{a} = \wh{b}$ as a shorthand for the
constraint $(\wh{a} = d_1 \land \wh{b} = d_1) \lor \ldots \lor (\wh{a} = d_n
\land \wh{b} = d_n),$ with $\Data = \set{d_1,\ldots,d_n}$.

\begin{definition}[Constraint Automaton~\cite{reo:ca}]
\label{def:ca}
	A \emph{constraint automaton} (over the finite data domain \Data) is a tuple
	$\mathscr{A}=\ltpl Q,\Names,$ $\to,Q_0 \rtpl$, where $Q$ is a set of
	states, $\Names$ is a finite set of port names, $\to$ is a
	subset of $Q\times \PS{\mkern2mu \Names}\times DC_{\Names}\times Q$, called
	the transition relation of $\mathscr{A}$, and $Q_0\subseteq Q$ is the set of
	initial states.
\end{definition}

\noindent
	We write $q\goesby{X|g}p$ instead of
	$(q,X,g,p)\in{\to}$.
	For every transition $q\goesby{X|g}p$, we require that $g$, the guard,%
	\index{guard!of constraint automata} is a
	$DC_X$-constraint. For every state $q\in Q$, there is a transition
	$q\goesby{\emptyset|\true}q$.

We define $\cas \subseteq \PS{\mkern2mu \Names}\times DC_{\Names}$ to be the set of
solutions for
all possible labels of the transitions of
\vspace*{-1mm}
constraint automata. That is, $X|g \in \cas$ if $X=\set{x_1,\ldots,x_n}$, $g = \bigwedge \wh{x_i}=v_i$, where $v_i \in \Data$, and there is a
transition $q\goesby{X|g'}q'$ such that $g$ satisfies $g'$. 
We call each $s
\in \cas$ a constraint automaton step.
Firing a
transition $q\goesby{X|g}p$ is interpreted as having dataflow at all
the ports in $X$, while excluding flow at ports in $\Names\setminus X$, when
the automaton is in the state $q$. The data flowing through the ports~$X$ must
satisfy the constraint $g$, and the automaton evolves to the state $p$.
\myref{fig}{caexamples} exemplifies the constraint automata for three \reo channels.
We do not define here the composition of constraint automata, but encode labels of constraint automata as labels of behavioural automata, whose composition has been defined in \myref{sec}{bacomp}.

\begin{figure}
  \centering
  \begin{tabular}{c@{~~}c@{~~}c@{~~}c@{~~}c}
  \wrap{\begin{tikzpicture}%
    [font=\small,inner sep=5pt,
     shorten >=1pt,>=stealth',node distance=2cm,auto,initial text=]
    \node[circle,draw=black,initial,initial where=above] (q) {$q$}; 
    \path[->] (q) edge [loop right] node {$ab\mysep \true$} ();
  \end{tikzpicture}}
  & ~~~~~~~~~~ &
  \wrap{\begin{tikzpicture}%
    [font=\small,inner sep=5pt,
     shorten >=1pt,>=stealth',node distance=2cm,auto,initial text=]
    \node[circle,draw=black,initial,initial where=above] (q) {$q$}; 
    \path[->] (q) edge [loop left] node {$a\mysep \true$} ()
                  edge [loop right] node {$ab\mysep \wh{a}=\wh{b}$} ();
  \end{tikzpicture}}
  & ~~~~~~~ &
  \wrap{\begin{tikzpicture}%
    [font=\small,inner sep=3pt,
     shorten >=1pt,>=stealth',node distance=30mm,bend angle=10,auto,initial text=,
     eliptic/.style={rectangle,rounded corners=2.5mm,draw=black,minimum height=6mm}]
    \node[eliptic,initial,initial where=above] (e) {$\mathtt{empty}$};
    \node[eliptic,right of=e] (f) {$\mathtt{full}(d)$};
    \path[->] (e) edge [bend left] node {$a \mysep \wh{a}=d$} (f)
              (f) edge [bend left] node {$b \mysep \wh{b}=d$} (e);
  \end{tikzpicture}}
  \end{tabular}
  \caption{From left to right, constraint automata for the \sdrain, \lossy and \fifo channels.}
  \label{fig:caexamples}
\end{figure}
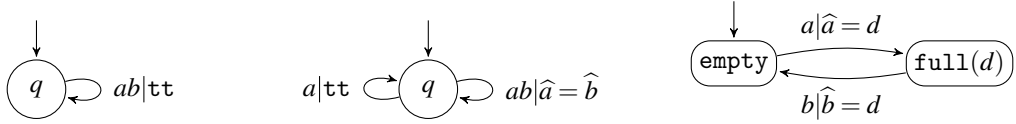

\subsection{Constraint automata as behavioural automata}
\label{sec:caasbd}

The \CA model assumes a finite data domain \Data, and that data constraints
such as $\true$, $\wh{a}\neq d$, or $\wh{a}=\wh{b}$ stand for simpler data
constraints that use $\wh{a} = d$ and the operators $\land$ and $\lor$.

The encoding of the constraint automaton $\dvar{A}=\tpl{Q,\Names,\to_\CA,Q_0}$
is the behavioural automaton
\vspace*{-2mm}
\[\encd{\dvar{A}\mkern2mu}_\CA = \tpl{Q,\lb[\Names],\to_\ba,\Obs}
\vspace*{-2mm}
\]
with $\lb[\Names]$, $\to_\bd$, $\Obs$, and the composition of labels defined as follows:
\begin{itemize}
  \item $\lb = \cas$, and $\alpha$ is defined as:
  $\alpha(X|\mwrap{\bigwedge_{i=1}^n \wh{x_i}=d_i}) ~=~ \tpl{\Names,X,\emptyset, X, \set{x_i\mapsto d_i}_{i=1}^n}.
  $
  
  \item We have $q \goesby{X|g}_\bd q'$ for $X|g \in \lb[\Names]$ if
  $q \goesby{X|g'}_\CA q'$ and $g$ satisfies $g'$.  

  \item Let $\mi{cas}_i = X_i|g_i$ be a solution for a label in a constraint automaton with ports $\Names_i$, for $i \in 1..2$. Then
  \vspace{-2mm}
  \[
  \scomp{\mi{cas}_1}{\mi{cas}_2} = \left\{\begin{array}{ll@{}}
  \multicolumn{1}{l}{(X_1 \cup X_2)|(g_1 \land g_2)}
   & \text{if~}
   X_1 \cap \Names_2 = X_2 \cap \Names_1 ~\land~ g_1 \compat g_2
  \\
  \bot & \text{otherwise}
  \end{array}\right.\]  
  where $g_1\compat g_2$ if for any   port $x \in X_1 \cap X_2$ and for any $d\in\Data$, $x=d$ satisfies $g_1$ iff $x=d$ satisfies $g_2$.

  \item $\Obs(q) = \excl{\Names}$ for every $q\in Q$. Recall that
  $\excl{\Names}=\{\ell ~|~ \alpha(\ell) = \ltpl P,F,\IP,\OP,$ $\data \rtpl, P_0 \cap F \neq
  \emptyset\}$. 
\end{itemize}

\begin{example} \label{ex:ca2bd}

Let $\dvar{A}_L=\tuple{Q_L,\dvar{N}_L,\to_1,Q_{1}}$ and 
$\dvar{A}_F=\tuple{Q_F,\dvar{N}_F,\to_2,Q_{2}}$ be the constraint automata of the \lossy and the \fifo channels, depicted in \myref{fig}{caexamples}.
The encoding of $\dvar{A}_L$ into behavioural automata is
$\tpl{Q_L,\lb[\dvar{N}_L],\to_{L},\Obs_L}$, depicted in the left hand side of \myref{fig}{cabacomposition},
where:\\
$~~Q_L =  \set{q}$,
$\dvar{N}_L = \set{a,b}$,
$\Obs_L(q) = \excl{\dvar{N}_L} \textrm{ for }q\in Q_L$,
$s_1(v) =
   ab|(\wh{a}=v\land\wh{b}=v)$,
$s_2(v) =   {a|(\wh{a}=v)}$,
and\\
$~\to_L =
   \set{\tpl{q,s_1(v), q} ~|~ v \in \Data} \cup
  \set{\tpl{q,s_2(v), q} ~|~ v \in \Data}$.

Similarly, the encoding of $\dvar{A}_F$ into behavioural automata is
$\tpl{Q_F,\lb[\dvar{N}_F],\to_{F},\Obs_F}$, also depicted in \myref{fig}{cabacomposition},
where:\\
$~~Q_F = \set{\tt empty} \cup \set{{\tt full}(v) ~|~ v \in \Data}$,
$\Obs_F(q) = \excl{\dvar{N}_F} \textrm{ for }q\in Q_F$,
$\dvar{N}_F  = \set{b,c}$,
$s_3(v) = {b|(\wh{b}=v)}$,
$s_4(v) = c|(\wh{c}=v)$,
~and~
$\to_F =
   \set{\tpl{{\tt empty},s_3(v), {\tt full}(v)} ~|~ v \in \Data} \cup
  \set{\tpl{{\tt full}(v), s_4(v), {\tt empty}} ~|~ v \in \Data}$.

The composed automaton $\encd{\dvar{A}_L}_\CA \bowtie
\encd{\dvar{A}_F}_\CA$ is depicted in the right hand side of \myref{fig}{cabacomposition}, where 
$
\scomp{s_1(v)}{s_3(v)} ~=~
ab|(\wh{a}=v\land \wh{b}=v)
\text{ ~~and~~}
\scomp{s_2(w)}{s_4(v)} ~=~
ac|(\wh{a}=w\land \wh{c}=v).
$
\end{example}

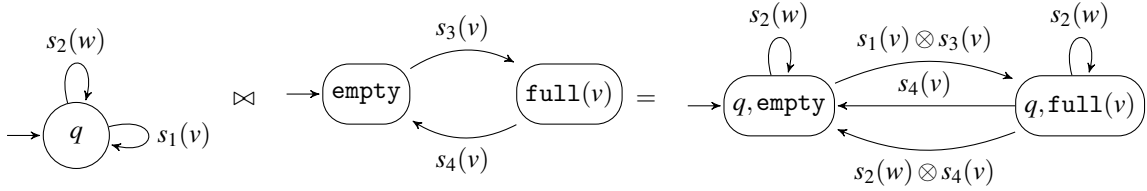
\begin{figure}
  \centering\small

  \begin{tabular}{c@{~~}c@{~~}c@{~~}c@{~~}c}
  \wrap{\begin{tikzpicture}%
    [shorten >=1pt,>=stealth',node distance=2cm,auto,initial text=]
    \node[state,initial] (q) {$q$}; 
    \path[->] (q) edge [loop above] node {$s_2(w)$} ()
                  edge [loop right] node {$s_1(v)$} ();
  \end{tikzpicture}}
  & $\bowtie$ &
  \wrap{\begin{tikzpicture}%
    [shorten >=1pt,>=stealth',node distance=27mm,auto,initial text=,
     eliptic/.style={rectangle,rounded corners=3.5mm,draw=black,minimum height=8mm}]
    \node[eliptic,initial] (e) {$\mathtt{empty}$};
    \node[eliptic,right of=e] (f) {$\mathtt{full}(v)$};
    \path[->] (e) edge [bend left] node {$s_3(v)$} (f)
              (f) edge [bend left] node {$s_4(v)$} (e);
  \end{tikzpicture}}
  & $=$ &
  \wrap{\begin{tikzpicture}%
    [shorten >=1pt,>=stealth',node distance=4cm,auto,initial text=,
     bend angle=22, 
     eliptic/.style={rectangle,rounded corners=3.5mm,draw=black,minimum height=8mm}]
    \node[initial,state,eliptic] (qe) {$q,\mathtt{empty}$};
    \node[eliptic,right of=qe] (qf) {$q,\mathtt{full}(v)$};
    \path[->] (qe) edge [loop above] node {$s_2(w)$} ()
              (qf) edge [loop above] node {$s_2(w)$} ()
              (qe) edge [bend left] node {$\scomp{s_1(v)}{s_3(v)}$} (qf)
              (qf) edge [swap] node[inner sep=1mm] {$s_4(v)$} (qe)
              (qf) edge [bend left] node {$\scomp{s_2(w)}{s_4(v)}$} (qe);
  \end{tikzpicture}}
  \end{tabular}
  \caption{Composition of $\encd{\dvar{A}_L}_\CA$ and $\encd{\dvar{A}_F}_\CA$, for any $v,w\in \Data$.}
  \label{fig:cabacomposition}
\end{figure}

The composed automata presented in \myref{ex}{ca2bd}, which differs from the lossy-FIFO, is equivalent to the product of the two associated constraint automata~\cite{reo:ca}, with respect to the atomic steps of the labels of the automata. We expect this equivalence to hold in general, but we do not give a formal proof here.
\JP{new text}

\section{Encoding Linda}
\label{sec:linda}

Linda, introduced by Gelernter~\cite{linda}, is seen by many as the first
coordination language. We describe it using Linda-calculus~\cite{linda:calculus}, and show how it can be modelled using behavioural automata.
Linda is based on the \emph{generative communication}
paradigm, which describes how different processes in a distributed environment
exchange data. In Linda, data objects are referred to as \emph{tuples}, and
multiple processes can communicate using a \emph{shared tuple-space},
where they can write or read tuples.

Communication between processes and the tuple-space is done by actions
executed by processes over the tuple-space. In general, these actions can
occur only atomically, that is, the shared tuple-space can accept and
execute an action from only one of the processes at a time. There are four possible
actions, $\mathbf{out}(t)$, $\mathbf{in}(s)$, $\mathbf{rd}(s)$, and
$\mathbf{eval}(P)$.
The actions $\mathbf{out}(t)$ and $\mathbf{in}(s)$ write and take values to and from the shared tuple-space, respectively. 
The action $\mathbf{rd}(s)$ is similar to $\mathbf{in}(s)$, except
  that the tuple $t$ is not removed from the tuple-space.
Finally,
$\mathbf{eval}(P)$ denotes the creation of a new process $P$ that
will run in parallel. We do not address $\mathbf{eval}(P)$ here
because it is regarded as a reconfiguration of the system.

\subsection{Linda-Calculus}
\label{sec:lindacalc}
We use the Linda-Calculus model, described by Goubault~\cite{linda:semantics}, to
give a formal description of Linda, studied also by
Ciancarini~\textit{et al.}~\cite{linda:calculus} and others. The Linda-Calculus
abstracts away from the local behaviour of processes, and focuses on the
communication primitives between a \emph{store} and a set of \emph{processes}.
Processes~$P$ are generated by the following grammar.
\begin{align}
   P &::= \mi{Act}.P \bigbar X \bigbar \mathbf{rec}X.P \bigbar P\lindaor P 
    \bigbar \mathbf{end}
   \\
   Act &::= \mathbf{out}(t) \bigbar \mathbf{in}(s) \bigbar \mathbf{rd}(s) 
\end{align}

\index{Linda!process}
\index{Linda!store}
\index{Linda!action ($a \in \Act$)}
We denote the set of all Linda terms as \linda.
The first case $\mi{Act}.P$ represents the execution of a Linda action.
The productions $X$ and $\mathbf{rec}X.P$ are used to
model recursive processes, where $X$ ranges over a set of variables, and $P \lindaor P$ is used to model local
non-deterministic choice. We assume that processes do not have free
variables, \ie, every~$X$ is bound by a corresponding $\mathbf{rec} X$.
Finally $\mathbf{end}$ represents termination.

We model a Linda store as a multi-set of tuples from a global set $\Tuple$.
Each tuple consists of a sequence of parameters, which can be
either a data value $v$ from a domain \Data (an actual parameter), or a
variable $X$ (a formal parameter).
We use the
$\oplus$~operator to denote multi-set construction and multi-set union. For
example, we write $M = t \oplus t = \mset{t,t}$ and $M \oplus M =
\mset{t,t,t,t}$, where $t$ is a tuple and $\mset{s,t}$ denotes a multi-set with the elements $s$ and $t$.

A \emph{tuple-space term} $M$ is a multi-set of processes and tuples,
generated by the grammar $M ::= P ~|~ t ~|~ M\oplus M$.
We adopt the approach of Goubault and
provide 
a set of inference rules that give the operational semantics of
Linda-Calculus.
A relation $\mi{match} \subseteq \Tuple \times \Tuple$
represents
the matching of two tuples. $(s,t) \in \mi{match}$ if $t$
has only \Data values, and there is a substitution $\gamma$ whose domain is
the set of free variables of $s$, such that $t = s[\gamma]$.
$u[\gamma]$ denotes the tuple or process $u$ after
replacing its free
variables according to $\gamma$. We also write $\gamma =
P/x$ to denote the substitution of $x$ by the process $P$, and
say $t$ $\gamma$-matches $s$ when $t$ matches $s$ and $t=s[\gamma]$.

\index{Linda!interleaved transition system}
\begin{definition}[Semantics of Linda]
The semantics of Linda is defined by the inference rules below.
\vspace*{-3mm}

\noindent
\hspace{-11mm}
\begin{minipage}{.43\textwidth}
\begin{align}
\multicolumn{2}{l}{$\mwrap{
  M \oplus P[\mathbf{rec} X.P/X] \longrightarrow M \oplus  P'
  \\\hline
  M \oplus \mathbf{rec} X.P \longrightarrow M \oplus  P'
  }$}
  \tag{rec}
\\
M \oplus P \lindaor P' & \longrightarrow M \oplus P
  \tag{left}
\\
M \oplus P \lindaor P' & \longrightarrow M \oplus P'
  \tag{right}
\end{align}
\end{minipage}
~
\begin{minipage}{.55\textwidth}
\begin{align}
M \oplus \mathbf{out}(t).P & \longrightarrow M \oplus P \oplus t 
  \tag{out}
\\
M \oplus \mathbf{rd}(s).P \oplus t & \longrightarrow M \oplus P[\gamma] \oplus t
  \textrm{~~if } t ~\gamma\textit{-matches } s
  \tag{rd}
\\
M \oplus \mathbf{in}(s).P \oplus t & \longrightarrow M \oplus P[\gamma]
  \phantom{\oplus t~\,}
  \textrm{~~if } t ~\gamma\textit{-matches } s
  \tag{in}
\\
M \oplus \mathbf{end} & \longrightarrow M
  \tag{end}
\end{align}
\end{minipage}
\end{definition}
\vspace{-1mm}

\begin{example}
\label{ex:lindaex}
The following sequence of transitions
illustrates the sending of data between two processes.
The labels on the arrows contain the names of the rules applied in each 
transition of Linda-Calculus. 
We use the notation $P(x)$ as syntactic sugar to denote a process $P$ where the variable
$x$ occurs freely.
\renewcommand{\arraystretch}{1.1}
\[\begin{array}{@{}c@{~}l@{~}c@{~}l@{}}
&
\multicolumn{3}{l@{}}{\mathbf{rd}(42,x).P(x) \oplus \mathbf{out}(42,43).\mathbf{end} \oplus \mathbf{in}(42,x).P'(x)}
\\
\goesby{(out)}&
\mathbf{rd}(42,x).P(x) \oplus \mathbf{end} \oplus \mathbf{in}(42,x).P'(x) \oplus \tpl{42,43}
&
\goesby{(end)}&
\mathbf{rd}(42,x).P(x) \oplus \mathbf{in}(42,x).P'(x) \oplus \tpl{42,43}
\\
\goesby{(rd)}&
P(43) \oplus \mathbf{in}(42,x).P'(x) \oplus \tpl{42,43}
&
\goesby{(in)}&
P(43) \oplus P'(43)
\end{array}\]
\renewcommand{\arraystretch}{1}
\end{example}

\subsection{Linda-calculus as behavioural automata}

We define an encoding function $\encd{\cdot}_\linda: \linda \to \ba$, from Linda tuple-space terms to behavioural automata. Furthermore, we
define the composition of atomic steps that preserve this semantics.
We encode each Linda process $P$ as a behavioural automaton,
and we create a special behavioural automaton that describes the multi-set of
available tuples.

Let $\overline{\mi{Act}} = \set{\overline{a} ~|~ a \in \mi{Act}}$ and
$\tAct = \set{\tau_a ~|~ a \in \mi{Act}}$.
A port $\overline{a}$ is regarded as a dual port of $a$, and 
flow of data on a port $\tau_a$ represents the flow on the ports $a$ and $\overline{a}$ simultaneously.
The intuition is that the encoding of processes yields behavioural automata
whose ports are actions in $\mi{Act}$; the encoding of tuples yield
behavioural automata whose ports are \emph{dual} actions in $\mi{\overline{Act}}$;
and the composition forces actions and dual actions to synchronise, \ie, to
occur simultaneously.
We define the
global set of ports to be $\Port = \mi{Act} \cup \overline{\mi{Act}} \cup
\tAct$, and define $\overline{\overline{a}}=a$.

Let $M = P_1 \oplus \cdots \oplus P_n \oplus T$  be a tuple-space term. In turn, let
$T = t_1 \oplus \cdots \oplus t_m$ and $m\geq 0$. We define the encoding of
$M$ into a behavioural automaton as follows.
\vspace*{-1.5mm}
\begin{equation*}
\encd{M}_\linda = \encd{P_1}_\linda \bowtie \cdots \bowtie 
\encd{P_n}_\linda \bowtie \encd{T}_\linda
\end{equation*}
\vspace*{-5.5mm}

\noindent
Hence, encoding $M$ boils down to encoding  Linda processes $P_i$ and the
Linda tuple-space $T$ into different behavioural automaton.
In both encodings of components and Linda tuple-spaces we define labels~$\lb$ as ports, that is, $\lb = \Port = \Act \cup \DAct \cup \tAct$, and its encoding as atomic steps by the function $\alpha$ defined below.
\vspace*{-4mm}
\begin{align*}
  \alpha(a) = \left\{\begin{array}{ll}
    \tpl{\Port,\set{a,\tau_{\mi{act}}},\emptyset,\emptyset,\emptyset} 
      & \text{if } a\in \Act\cup \DAct, \set{\mi{act}} = \set{a,\overline{a}}\cap\Act
    \\
    \tpl{\Port,\set{a},\emptyset,\emptyset,\emptyset} 
      & \text{if } a\in \tAct
   \end{array}\right.
\end{align*}

\noindent
The composition of two labels $a_1,a_2 \in \lb$ is defined as follows.
\vspace*{-2mm}
\[
  \scomp{a_1}{a_2} = \left\{\begin{array}{ll}
  \tau_{\mi{act}}
    &\text{if~~}
  \mi{a}_1 \notin \tAct ~\land~ 
    \mi{a}_2 \notin \tAct ~\land~ 
    \mi{a}_1 = \overline{\mi{a}}_2
  \\
  \bot & \text{otherwise,}
  \end{array}\right.
\]
where $\set{\mi{act}} = \set{a_1,a_2}\cap \Act$.
The tuple-space is used to enforce every action $a$ performed by a process to
synchronise with the corresponding action $\overline{a}$ in the tuple-space
encoded as a behavioural automaton. The definition of $\scomp{}{}$ replaces
every pair of ports with dataflow $a$ and $\overline{a}$ by a new port with
dataflow in $\tau_{\mi{act}}$.

We encode a Linda process $P$ as 
$\encd{P}_\linda =  \tpl{Q_P, \lb, \to_P, \Obs}$,
with components as defined below.
\begin{itemize}
  \item The set of states $Q_P$ is given by $Q_P = \mi{reach}(P)$, where
          \[ \begin{array}{rcl}
          \mi{reach}(\mathbf{out}(t).P) &=& \set{\mathbf{out}(t).P} \cup \mi{reach}(P)
          \\
          \mi{reach}(\mathbf{rd}(s).P) &=& \set{\mathbf{rd}(t).P} \cup
          (\bigcup\set{\mi{reach}(P[\gamma]) ~|~ s~\gamma\textit{-matches}~t})          
          \\
          \mi{reach}(\mathbf{in}(s).P) &=& \set{\mathbf{in}(t).P} \cup 
          (\bigcup\set{\mi{reach}(P[\gamma]) ~|~ s~\gamma\textit{-matches}~t})          
          \\
          \mi{reach}(P \lindaor P') &=& \set{P \lindaor P'} \cup \mi{reach}(P) \cup \mi{reach}(P')
          \\
          \mi{reach}(\mathbf{end}) &=& \set{\mathbf{end}}
          \end{array}\]
          
  \item The transition relation $\to_P$ is given by the following conditions.
    \renewcommand{\arraystretch}{0.8}
    \[\begin{array}{r@{~}c@{~}lcl@{~~~~~~~~~~~~~~~~~~~~~~~~~~~}r@{~}c@{~}lcl}
     \mathbf{out}(t).P' & \goesby{
        \mathbf{out}(t)}
        & P'
        &\mi{if}&
        t \in \Tuple
     &
     P_1\lindaor P_2 & \goesby{~~s~~}
        &P'_1
        &\mi{if}&
        P_1 \goesby{s} P'_1
     \\
     \mathbf{rd}(s).P' & \goesby{
         \mathbf{rd}(t)}
         & P'[\gamma]
        &\mi{if}&
        s~\gamma\textit{-matches}~t
     &
     P_1\lindaor P_2 & \goesby{~~s~~}
        & P'_2
        &\mi{if}&
        P_2 \goesby{s} P'_2
     \\
     \mathbf{in}(s).P' & \goesby{
         \mathbf{in}(t)}
         & P'[\gamma]
        &\mi{if}&
        s~\gamma\textit{-matches}~t
    \end{array}\]
    \renewcommand{\arraystretch}{1}

  \item $\Obs(q) = \Act \cup \DAct$ for every state $q$.
\end{itemize}

\medskip

\noindent
We now encode a Linda tuple-space $T$ as
$\encd{T}_\linda = \tpl{Q_T, \lb, \to_T, \Obs}
$
with components as defined below.
\begin{itemize}
  \item $Q_T = \PS{\MS(\Tuple)}$, where $\MS(X)$ is a multi-set over the set $X$.
  
  \item The transition relation $\to_T$ is given by the following conditions:
\\
$M \goesby{ \overline{\mathbf{out}(t)}}
  M \oplus t \textit{ if }
  t \in \Tuple$,~~~
$t \oplus M \goesby{ \overline{\mathbf{rd}(s)}}
  t \oplus M \textit{ if }
  s~\mi{matches}~t$, and~~~
$t \oplus M \goesby{ \overline{\mathbf{in}(s)}}
  M \textit{ if }
  s~\mi{matches}~t$.

  \item $\Obs(q) = \Act \cup \DAct$ for every state $q$, as in the encoding of Linda processes.

\end{itemize}

Note that the input and output ports of the atomic steps obtained with $\alpha$, introduced in \myref{sec}{atomicstep}, are always the empty
set, that is, the data value flowing through the ports is not relevant, since the
name of the port uniquely identifies the data. Alternative approaches to
implement the encoding into behavioural automata that use the data values are
also possible, but less transparent.
\JP{dropped comment about compositionality.}

\begin{example}

Recall the example presented in the end of \myref{sec}{lindacalc} of a sequence of transitions of a tuple-space term in
Linda-Calculus. We present below a simplified version of this example.
\[  \mathbf{rd}(42,x).P(x) \oplus \mathbf{out}(42,43).P'
~~~\goesby{(out)}~~~
\mathbf{rd}(42,x).P(x) \oplus P' \oplus \tpl{42,43}
~~~\goesby{(rd)}~~~
P(43) \oplus P' \oplus \tpl{42,43}
\]
The
corresponding transitions in the encoded behavioural automaton are presented
below.
\[\begin{array}{l@{~~}c@{~~}l}
 \encd{\mathbf{rd}(42,x).P(x)}_\linda \bowtie \encd{\mathbf{out}(42,43).P'}_\linda \bowtie \encd{\emptyset}_\linda
 &\goesby{\tau_{\mathbf{out}(42,43)}}\\
 \encd{\mathbf{rd}(42,x).P(x)}_\linda \bowtie \encd{P'} \bowtie \encd{\tpl{42,43}}
 &
 \goesby{\tau_{\mathbf{rd}(42,43)}}&
  \encd{P(43)}_\linda \bowtie \encd{P'} \bowtie \encd{\tpl{42,43}}
\end{array}\]
\end{example}

Observe that we assume
an initial empty tuple-space, which is encoded as $\encd{\emptyset}_\linda$.
A more
careful analysis shows a one-to-one correspondence between the traces of the Linda-calculus term and the traces of the behavioural
automaton, which we do not elaborate in this paper.


\section{Exploiting concurrency predicates}
\label{sec:cp}

We introduced a unified model for synchronous coordination that explicitly
mentions concurrency predicates, which indicate which actions 
require synchronisation.
We now exploit more complex definitions of concurrent predicates for \reo and Linda than in our previous examples, and briefly describe a practical application of behavioural automata in 
a distributed framework.

\subsection{Complex concurrency predicates}

In our examples concurrency predicates of \reo hold when some shared ports from a
composed automaton have dataflow (\myref{eq}{excl}), and concurrency
predicates of Linda allow only a special set of actions \tAct to run concurrently. We now present other concurrency predicates that capture notions such as context dependency and priority.

\myparagraph{Reo}
Other semantic models for \reo, such as connector colouring~\cite{reo:cc} and
\reo automata~\cite{reo:ra}, capture the notion of \emph{context dependency},
a feature missing in constraint automata. By modelling context dependency we
avoid the undesired behaviour of the composed connector in
\myref{fig}{cabacomposition} where data is lost when the \fifo buffer is
empty, represented by the label $s_2(w)$.

To avoid data from being lost, we replace the \lossy channel by
a context dependent \lossy channel, which is built based on the \lossy channel
by replacing the label $s_2(w)$ by a label $s_2^b(w)$.
This new label has the
same atomic step, \ie, $\alpha(s_2(w)) = \alpha(s_2^b(w))$, but can be
executed in parallel only if its neighbours require the port $b$ to have no dataflow.
This condition
is enforced by adapting the definition of
concurrency predicates to check wether a given set of ports $Y$ requires synchronisation. 
\vspace*{-1mm}
\begin{align}
\mi{cp}_{\mi{ctx}}(P_0,Y) = \set{s^X ~|~ s^X \in \excl{P_0} ~~\lor~~ X \cap Y \neq \emptyset}
\vspace*{-2mm}
\end{align}
In our example, we avoid the losing of data by defining $\Obs(q) =
\mi{cp}_{\mi{ctx}}(ab,\emptyset)$, $\Obs(\mathtt{empty}) =
\mi{cp}_{\mi{ctx}}(bc,b)$, and $\Obs(\mathtt{full}(v))
=\mi{cp}_{\mi{ctx}}(ab,c)$. The label $s_2^{b}(w)$ 
is in $\Obs(\mathtt{empty})$
but not in $\Obs(\mathtt{full})$, \JP{we say in the beginning that we omit restriction when clear from the context.} \ie, $s_2^b(w)$ can be performed
independently of the \fifo channel only when the \fifo is full.
Other important details, such as the composition of labels of the form $s^X$, are not presented in this paper. A more precise and complete formulation can be found in Proen\c{c}a's Ph.D. thesis (Sections 3.6.2 and 4.4.2 of~\cite{proenca:phd}).

\myparagraph{Linda}
Consider now that Linda processes have a total order $\preceq$, representing a ranking among processes. When two processes can interact simultaneously with the shared tuple-space, only the higher rank should be chosen. We present only a sketch of this approach due to space limitation.

We start by tagging labels $\ell$ of the Linda behavioural automata with the process that executes it.
For example, a label $\ell$ of an automaton of a process $p$ is renamed to
$\ell^p$. Labels of the shared tuple-space are not changed. The composition
of labels must be such that
$\scomp{\ell^p}{\overline{\ell}} = \tau_{\ell}^{p}$.
It is then enough to change the concurrency
predicates of the automata of each process $p$ in state $q$ to
$\Obs(q) =
\Act \cup \DAct \cup \set{\tau_{\ell}^{x} ~|~ \tau_{\ell} \in \tAct ~\land~ x \preceq p ~\land~ q \neq \mathbf{end}}
$
and leave the concurrency predicate of the automaton of the shared tuple-space unchanged. Hence,
a transition cannot be performed in parallel if it is in \Act or \DAct, or if
it is a $\tau$ action from a process with lower priority and the current
process is not yet stopped.

\subsection{Increased scalability via decoupled execution}
\label{sec:dreams}

We use the behavioural automata model in a distributed framework,
\dreams, where several independent threads run
concurrently~\cite{proenca:phd}.  Each thread has its own behavioural
automaton, and communicates only with those threads whose behavioural automata share ports with its own automata.
The details regarding this tool are out of the scope this paper, but we
explain how it benefits from using behavioural automata.

The diagram in \myref{fig}{dreams} depicts the configuration of a system in \dreams,
where each cloud represents an independent thread of execution, and edges
represent communication links between threads whose automata share ports. The direction of each edge only illustrates the expected direction of dataflow.
For efficiency reasons, and to allow a lightweight reconfiguration, \dreams
does not create the complete behavioural automaton of a connector. Instead, it collects   
only the behaviour of the current round.

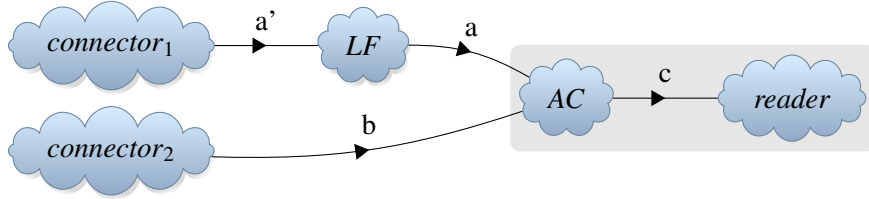
\begin{figure}
\centering
\begin{tikzpicture}[%
      text=black,reodist,node distance=33.5mm,
      vdist/.style={node distance=7mm},
      actor/.style={bcld,text=black,inner sep=1.5mm,cloud puffs=12},
      actor2/.style={bcld2,text=black,inner sep=1.5mm,cloud puffs=12},
      directed/.style={decoration={ markings, mark=at position .50 with {\arrow[black]{triangle 60}}},postaction={decorate}},
      lbl/.style={above,inner sep=2mm}
      ]
  \node[actor,inner sep=1mm] (c1) {\textit{connector}$_1$};
  \node[below of=c1,vdist] (m1) {};
  \node[right of=m1] (m2) {};
  \node[actor,inner sep=1mm,below of=m1,vdist,minimum height=1mm] (c2) {\textit{connector}$_2$};
  \node[actor,above of=m2,vdist,cloud puffs=9] (lf) {$\mathit{LF}$};
  \node[actor2,right of=m2, node distance=27mm,cloud puffs=9] (ac) {$\mathit{AC}$};
  \node[actor2,right of=ac, node distance=30mm] (r) {\textit{reader}};
  
  \draw[directed] (c1) -- node[lbl] {a'} (lf);
  \draw[directed,bend angle = 10, bend right] (c2) to node[lbl] {b} (ac);
  \draw[directed,bend angle = 15, bend left] (lf) to node[lbl] {a} (ac);
  \draw[directed] (ac) -- node[lbl] {c} (r);
  
  \reobox{(ac)(r)}
\end{tikzpicture}
\caption{Configuration of a system in \dreams.}
\label{fig:dreams}
\end{figure}

Knowing that only the labels of the automata relevant for the current round
are composed, and assuming that the locality property introduced
in~\myref{def}{localityba} holds, we can perform local steps that, as the name
suggests, involve only a subpart of the system. Recall the example of the
lossy alternator, presented in \myref{sec}{lossyalternator}. The diagram in
\myref{fig}{dreams} uses the same example, in a context where two arbitrary
large connectors $\mi{connector}_1$ and $\mi{connector}_2$  are attached to
the source of the lossy alternator, and a $\mi{reader}$ component is attached
to the sink of the lossy alternator. Consider that the $\mi{reader}$ can always
receive any data value, that is, its behavioural automaton has a single state,
and a transition labelled by $r(v)$ for every data value $v$, such that
$\alpha(r(v)) = \tpl{c,c,c,\emptyset,\set{c\mapsto v}}$.

Observe that we do not use explicitly the composed connector $\mi{LF}\bowtie
\mi{AC}$, but $\mi{LF}$ and $\mi{AC}$ as independent entities instead, since
the \dreams framework can postpone the composition of their labels to runtime.
Consider that the $\mi{AC}$ automaton is in state $q_1(v)$, hence it can
perform a step $s_2(v)$, writing a value $v$ to the port $c$. In this example
$\mi{AC}$ is connected via the ports $a$, $b$, and $c$. The
label $s_2(v)$ does not have dataflow on $a$ nor on $b$, and the reader can
perform a label $r(v)$ because $\scomp{s_2(v)}{r(v)}\neq \bot$. Using the
concurrency predicate in \myref{eq}{excl}, we conclude
that $\scomp{s_2(v)}{r(v)}$ is in the concurrency predicates of $\mi{LF}$ and
$\mi{connector}_2$. Furthermore, from the locality property we conclude that
all other connectors not attached to $\mi{AC}$ also allow
$\scomp{s_2(v)}{r(v)}$ to be executed concurrently. Hence, \dreams can chose to
perform this step by analysing only the behaviour of $\mi{AC}$ and
$\mi{reader}$, depicted by a grey box. 

The instantiations of Linda and Reo yield a similar result. The shared
tuple-space can communicate with a single process at a time, without
synchronising with every other process. \reo can, for example, send data from
a full \fifo independently of the behaviour of the connector attached to its
sink port. The benchmarks performed for the \dreams
framework~\cite{proenca:phd} show optimistic results regarding the use of
local steps in synchronous coordination.


\section{Conclusion}
\label{sec:conclusions}

We introduce behavioural automata to model coordination systems. The three
main concepts that underlie behavioural automata are \emph{atomicity},
\emph{composability}, and \emph{dataflow}. We allow a sequence of actions that
cannot be interleaved with interfering instructions (atomicity), we construct
more complex systems out of building blocks that can be analysed independently
(composability), and we represent the data values that are exchanged between
components (dataflow).

Behavioural automata unify existing dataflow-oriented models with synchronous constructs by leaving open the definitions of composition of labels and of concurrency predicates.
The focus of behavioural automata is on concurrent systems, and on avoiding synchronisation of actions whenever it is unnecessary. 
By capturing a multitude of coordination models, we allow any of these models to be included in implementations based on behavioural automata, such as the \dreams framework.

As future work, we expect to formally show the correctness of the encodings of Reo and Linda. We would also like to discover which properties can be shown for behavioural automata that are directly reflected on encoded models. 
A more practical track of our work involves the development of tools.
Further development of \dreams to make it ready for use by a broader community is in our agenda.

\reviewin{1: How can the product of n ($n>2$) automata be computed? Do the synchronization constructs change? Is it more complicated to deal with the interleaved labels?\\
Considering the example presented on Figure 1, is it possible to have the case when W1 and W2 share the port name, saying both use the "a" port to write data? How to deal with this situation if it is possible, and if a conflict occurs?\\
Considering the example presented on Figure 1, is it possible to have the case when W1 and W2 share the port name, saying both use the "a" port to write data? How to deal with this situation if it is possible, and if a conflict occurs?}
\JPin{Composition achieved by shared variables. That is how we compose $n>2$ automata. Need to epmhasise. I now mention that output ports replicate and input ports merge data.}

\reviewin{1: I am wondering if internal ($\tau$) behaviors can be covered by the
local actions used in this work. Adding some clarifications, in this respect,
would be interesting, being given the importance of such behaviors when
verifying concurrent systems.}
\JPin{did not really get it.}

\reviewin{1: Label matching depends on a compatibility condition, but how is it defined? For example, can a label m which receives a real be compatible with a label m which sends an integer?}
\JPin{no type information is used yet. It is the developers choice to define the compatibility relation, by defining when the composition relation is defined. I do not address this.}

\reviewin{1: As far as data-flow is concerned, on which semantics rely the data-dependencies? Is it late or early? I assume the authors implicitly use one of them, but this should be mentioned explicitly, cf. the paper "Modal logics for mobile process" by Milner et al.}
\JPin{In \dreams it seems that we assume late bisimilarity. When the individual behaviour of each actor is broadcasted, the input ports with yet unknown data are assumed to be a valid transition for any possible input data value. We now mention this.}

\reviewin{1: How is the definition of the product of behavioral automata related to synchronous product previously defined in the literature?}
\JPin{We already said that, wrt the CA, we expect to be the same, but we did not formally prove it. We now refer to the synchronous product from SCCS.}

\reviewin{2: The solution seems to be to make the concurrency predicate a function of the labels in the universe, so it typically expands when the label universe does, but this complicates matters considerably. Explaining this design choice would *really* be appreciated, because from all appearances the opposite choice of specifying the labels that are not concurrent and assuming that all labels not mentioned are always concurrent (thereby building in open world assumption) would be far more natural.
\\
~~~~~(...) the observation function O gives no hint that it is typically a function of the universe of ports P.}
\JPin{Main comment of Dave as well. Maybe define \Obs as a predicate states when a label requires synchronisation (the negation of the current definition). It is still global, but it gives the idea of describing local information. I applied this change, and I also require labels to be restricted.}

\reviewin{2: page 3: "data values" - typically it seems you are not specifying actual data values, but binding values to a variable that is used later. Some discussion might be nice.}
\JPin{This fits the ``grouping of labels" discussion from the thesis, that I do not include here. Maybe add a sentence starting with ``In our implementation, we ...". I do not address this comment, only briefly when referring to the late-bisimilarity.}

\reviewin{2: Motivate concurrency predicates by saying that you will define observation functions in terms of them. In fact, I don't understand why you need both concepts...?}
\JPin{They are the same, with and without references to states. Maybe some renaming can help? I now merged both concepts.}

\reviewin{3: the motivation behind coming up with behavioral automata remains a
mystery to me:
\\- is it the elegance of the formalism? 
\\- because no nice unified model of other languages existed so far?
\\- because they offer insight and suggest further development of the source
  languages?
\\- because they allow building analysis tools?
}
\JPin{I now emphasise in the introduction that Dreams (scalable and distributed implementation of \reo like languages) was the main motivation.}

\reviewin{3: Section 5.1 makes the usage of concurrency predicates appear
somewhat ad hoc or an afterthought. The notion of locality developed in 2.4 is
only briefly used once in Section 5. In general, it feels like Section 5 is
the most interesting (yet most unsatisfying) and would deserve to be developed
carefully and given a stronger emphasis within the paper. }

\reviewin{3,4: There is no discussion of related work.
\\
~~~~~4: how it relates to and advances the SoTA}
\JPin{Where to add related work, and what work to refer to? I only mention SCCS, but nothing more.}

\reviewin{4: it is difficult to understand how much the introduced formalism
is usable in practice. It might be useful to give an idea on the how the
formalism can be supported by a tool, which promotes and eases its adoption.}
\JPin{we mention dreams in section 5. We now also refer to it in the intro.}

\bibliographystyle{eptcs}
\bibliography{src/extras/bibliography}

\end{document}